\documentclass[10pt]{elsarticle}
\journal{Wave Motion}

\usepackage{graphicx}
\usepackage{amsmath,amsfonts,amsthm,bm,mathtools} 
\usepackage{lipsum}
\usepackage{float}
\usepackage{pgfplots}
\usepackage{graphicx,amsmath,upgreek,bm,amssymb}
\usepackage{subcaption}
\usepackage{siunitx}
\usepackage{makecell}
\usepackage{booktabs}
\usepackage{multicol}
\usepackage{xcolor,colortbl}
\usepackage[bb=boondox]{mathalpha}

\usepackage{geometry}
\usepackage{setspace}
\usepackage{longtable}
\usepackage[tableposition=top]{caption}
\usepackage{multirow}

\usepackage[font=small,
justification=justified,
format=plain]{caption} 
\newlength\fwidth
\usepackage{natbib}
\definecolor{mycite}{gray}{0.0}

\usepackage[colorlinks=blue]{hyperref}
\hypersetup{
	colorlinks=false,
	pdfborder={0 0 0},
}
\hypersetup{
	colorlinks=true,
	linkcolor=blue,
	citecolor= blue,
	filecolor=magenta,      
	urlcolor=blue,
}

\definecolor{rev1}{rgb}{0.8,0,0}

\usepackage[misc]{ifsym}
\usepackage{fontawesome}
\usepackage{ragged2e}

\usepackage{todonotes}

\usepackage{lineno}

\DeclareMathOperator*{\argmin}{arg\,min}

\begin{document}
	
	\begin{frontmatter}
		\title{Physics-Informed Machine Learning for the Inverse Design of Wave Scattering Clusters}	
		\author[add1]{Joshua R.~Tempelman\corref{cor1}}
		\cortext[cor1]{Corresponding author}
		\ead{jrt7@illinois.edu}
		\author[add2]{Tobias Weidemann}
		\ead{tobias.weidemann@ila.uni-stuttgart.de}
		\author[add3]{Eric B.~Flynn}
		\ead{eflynn@lanl.gov}
		\author[add1]{Kathryn H.~Matlack}
		\ead{kmatlack@illinois.edu}
		\author[add1]{Alexander F.~Vakakis}
		\ead{avakakis@illinois.edu}
		\address[add1]{Department of Mechanical Science and Engineering, University of Illinois at                Urbana-Champaign, Urbana, IL, USA}
		\address[add2]{Institute of Aircraft Propulsion Systems, University of Stuttgart, Germany}
		\address[add3]{Space Remote Sensing and Data Science, Los Alamos National Laboratory, Los Alamos, NM, USA}

		\begin{abstract}
				Clusters of wave-scattering oscillators offer the ability to passively control   wave energy in elastic continua. 
				However, designing such clusters to achieve a desired wave energy pattern is a highly nontrivial task.
				While the forward scattering problem may be readily analyzed, the inverse problem is very challenging as it is ill-posed, high-dimensional, and known to admit non-unique solutions. 
				Therefore, the inverse design of multiple scattering fields and remote sensing of scattering elements remains a topic of great interest.
				Motivated by recent advances in physics-informed machine learning, we develop a deep neural network that is capable of predicting the locations of scatterers by evaluating the patterns of a target wavefield.
				We present a modeling and training formulation to optimize the multi-functional nature of our network in the context of inverse design, remote sensing, and wavefield engineering.
				Namely, we develop a multi-stage training routine with customized physics-based loss functions to optimize models to detect the locations of scatterers and predict cluster configurations that are physically consistent with the target wavefield.
				We demonstrate the efficacy of our model as a remote sensing and inverse design tool for three scattering problem types, and we subsequently applicability for designing clusters that direct waves along preferred paths or localize wave energy.
				Hence, we present an effective model for multiple scattering inverse design which may have diverse applications such as wavefield imaging or passive wave energy control.
		\end{abstract}
		
		
	\end{frontmatter}
	
	\newcommand{\var}{\bm{x}}
	\newcommand{\sol}{\psi}
	\newcommand{\domain}{\Omega}
	\newcommand{\scatterers}{\mathcal{X}}
	\newcommand{\iter}{\gamma}
	\newcommand{\paramvec}{\bm{\theta}}
	\newcommand{\mean}[2]{\mathrm{mean}_{#1}\left\{#2\right\}}
	\newcommand{\card}[1]{\mathrm{card}\left(#1\right)}
	\newcommand{\ie}{i.\,e., }
	\newcommand{\mse}{\mathrm{MSE}}
	\newcommand{\ee}{\mathrm{e}}
	\thispagestyle{empty}

\section{Introduction}
\label{Sec:S1_Intro}

The capacity to control the scattering of wave energy in elastic continua has far-reaching applications such as 
energy harvesting~\cite{Zhou2021,Wei2017,Lee2024,Wang2021}, acoustic imaging~\cite{Ma2022,Liu2016,Parker,Assouar2018},  and wave steering~\cite{Deng2021,Torrent2013,Pu2022a}, to name a few.
To this end, local mechanical oscillators attached to the surface of an elastic continuum offer an effective tool for achieving this task~\cite{Merkel2015,Packo2019,Boutin2006,Norris1995}; 
such wave scattering mechanisms have led to the conception of practical periodic meta-beams~\cite{Wang2005,Jr.2019}, meta-plates~\cite{Assouar2012,Chaplain2020}, and meta-surfaces~\cite{Pu2020}, which can control wave propagation in the underlying continuum.
Moreover, the complicated interactions between the various scattering components of finite (non-periodic) surface oscillator clusters enables the suppression or redirection of surface~\cite{Pu2021,Pu2022} and seismic waves~\cite{Xu2023}.
Whereas forward modeling schemes are readily available to compute the effects of surface scatterers~\cite{Torrent2013,Pu2022},
the task of inverse designing scattering wave fields, that is, to compute a necessary scattering cluster to produce an observed wavefield, remains largely unexplored.

Due to the inherent complexity of wave scattering problems, the remote sensing or inverse design of surface scatterers is a highly nontrivial and ill-posed problem that admits non-unique solutions~\cite{Colton2019}.
While the assumption of periodic boundaries may greatly reduce the complexity of modeling surface scatterers~\cite{Yu2014}, such a framework requires the implementation of many unit cells in order to achieve effective periodicity in the clusters and is thus less attractive for practical wave scattering designs.
To this end, the method of multiple scattering offers an attractive framework for analyzing the effect of finite (non-homogenized) clusters of point-scatterers~\cite{Torrent2013}.
Moreover, such a modeling scheme is broadly applicable since the multiple-scattering formulations are general in the sense that they apply to any continua with a known Green's function~\cite{Pu2021,Pu2024}.
However, while the multiple scattering formulation allows for the convenient analysis of a known scattering cluster, inverse solutions of multiple scattering wavefields are not easily achieved; this is the focus of the current work.
From a practical perspective, the inverse solution offers abundant utility since it allows for the detection or design of scattering elements to recover an observed wavefield or achieve a desired wavefield, depending on the application.

Several recent studies have approached the multiple scattering inverse design problem from a variety of frameworks.
Packo \textit{et al.} presented an analytical approach to inverse design grated scattering fields based on periodically spaced scatterers in order to control wave transmission behavior~\cite{Packo2019}.
The same authors later contrived an analytical inverse optimization scheme whereby the effective force of the scatterer point load could be optimized to manipulate the form of the scattering solution~\cite{Packo2021}.
Moreover, related work has demonstrated that disordered periodic arrays of scatters can be designed to enable anomalous refraction and localization~\cite{Cao2020}.
Most recently, Capers proposed an adjoint-based approach to optimize the locations of pinned-points and mass loaded on a thin plate for the purpose of energy localization~\cite{Capers2023}.
Despite the advances of these past works, the solutions of traditional optimization must be reformulated for each wave field of interest, and thus a more generalized inverse modeling approach is still desired.

Recently, machine learning (ML) techniques have proven quite useful in the context of inverse design with applications ranging from
photonics~\cite{So2020,Liu2021}, mechanics~\cite{Zheng2023}, and acoustics~\cite{Bianco2019}, to name a few.
This surge is due in part to the recent prominence of automatic differentiation and GPU-based computing, both of which have drastically decreased the computational cost of implementing ML methods.
A key advantage of data-driven inverse models is their generality, since they can cheaply evaluate new targets once trained (so long as the new target is sufficiently within the training distribution)~\cite{Gurney2018}.
In general, most ML-based inverse design frameworks employ neural networks (NNs) to featurize (interpret) an inverse design target and then map it to a set of parameters~\cite{So2020}.
Broadly speaking, NNs perform as function approximators that map information from one domain to another by training the weights and biases parameterizing the NN to minimize a loss function~\cite{Hornik1989}.
Accordingly, NN-based architectures have been adapted to solve several challenging inverse design problems in the wave propagation community such as acoustic filter design
\cite{Cheng2022,Bacigalupo2019,Mahesh2021},  
metamaterial topology optimization~\cite{He2021,Du2023,Kollmann2020},
and phononic dispersion design~\cite{Long2019,He2023,Li2023}.

While traditional NN models rely solely on the observable data to learn the desired mapping, recent advances in  scientific ML (SciML) have presented novel techniques for incorporating domain-specific physics knowledge into ML frameworks~\cite{Cuomo2022}.
Perhaps most notably was the advent of \textit{physics-informed neural networks} (PINNs) which were originally developed to solve partial differential equations by embedding balance laws and boundary conditions directly into the loss function~\cite{Raissi2019}; similar methods have been proposed for inverse PDE solutions~\cite{Gao2022}.
Subsequently, PINN-inspired architectures have been proposed to evaluate a wide variety of physical phenomena including 
fluid dynamics~\cite{Choi2022,Cai2021}, photonics~\cite{Wiecha2021}, and solid mechanics~\cite{Haghighat2021}.
In the framework of wave mechanics, PINN-inspired networks have been employed for inverse design problems related to nano-optics~\cite{Chen2020}, metasurface design~\cite{Li2022}, and metamaterial design~\cite{Fang2020}.
Recently, a physics-constrained framework capable of detecting the geometry of 2-dimensional scatterers in acoustic wavefields was presented~\cite{Wu2022}; this work was subsequently extended to perform inverse designs for single 2D scattering sources~\cite{Nair2023}.

Motivated by the recent advances in SciML for inverse design, we set out to approach the inverse multiple-scattering problem from a physics-based deep learning framework. 
In particular, we seek to address the following practical design questions for multiple scattering formulations: 
	(i) \textit{Scatter detection}: Given a observed scattering field, can a model accurately predict the locations of the point scatterers responsible for the observed wavefield?
	(ii)  \textit{Inverse design}: Can the predicted scatterers be optimized to reconstruct the desired scattering field? We note this differs question (i) due to the expected non-uniquness of multiple scattering solutions.
	(iii)  \textit{Wavefield engineering}: Can the model produce scatterer clusters capable of approximating synthetically generated wavefield energy patterns?
To address questions (i)-(iii), we implement a convolutional auto-encodeder (CAE) coupled with a multi-layer perception (MLP) network which is capable of interpreting target wavefields and outputting a predicted scattering cluster.
Moreover, we engineer customized loss modules based on multiple scattering theory to encourage our network to recover physically meaningful results. 
Additionally, we adopt a customized multi-stage training strategy that allows our model to perform multi-functional goals such as scatterer location identification (remote sensing), multiple scattering wavefield reconstruction (inverse design), and cluster design for a targeted synthetic energy distribution (wavefield engineering).
We validate our approach by quantifying the inverse design accuracy across a multitude of multiple-scattering problem types, and demonstrate that the model is capable of producing the necessary scattering cluster designs for achieving synthetically engineered wavefield localization patterns.

Accordingly, the outline of this paper is as follows. Section~\ref{Sec:S2_Problem} outlines the mathematical details of the multiple-scattering forward solution and poses the inverse design problem statement for three distinct scattering problem classes.
Section~\ref{Sec:S3_ML} presents the machine learning framework utilized to solve the inverse problem, details the custom physics-informed loss functions, and describes the customized training routine and hyperparameter optimization scheme.
The results of our model are given in section~\ref{Sec:S4_Resuls} which details the training performance and demonstrates the  efficacy of the models for remote sensing, inverse design, and wavefield engineering.
Lastly, section~\ref{Sec:S5_Conclustions} offers concluding remarks and  a discussion of possible applications and future work.
\section{Multiple Scattering Problem Formulation}
\label{Sec:S2_Problem}

Our problem is posed around the method of multiple scattering. 
This method provides analytical solutions for the interactions of an incident harmonic wave with clusters of discrete point-scatterers~\cite{Pu2021}.
With the solution to the forward problem given, e.g., solving a wave scattering problem given some known scattering elements, we pose the question of inverse design and scattering field engineering.
In this regard, the inverse modeling goal is to solve for the set of scattering elements given some known wavefield pattern.
Accordingly, this section provides the mathematical framework for solving the forward problem, formulates the inverse design statement, and describes the three distinct categories of multiple scattering problems studied in this work.

\subsection{Mathematical Model}

We consider an infinite 2-dimensional elastic domain $\Omega\in\mathbb{R}^2$ satisfying the assumptions of  Kirchhoff-Love plate theory.
An incident harmonic force $f_0(t) =f_0e^{i\omega t}$ at location $\var_0$ produces an incident wave, and finite cluster of linear oscillators attached to a free boundary $\partial\Omega\in\Omega$ scatter the incident energy (Fig~\ref{FIG: ProbStatement}(a)). Herein, we consider the set
$
	\scatterers = \{\var_{\alpha},m_\alpha,c_{\alpha},k_{\alpha}: \ \alpha = 1,2,\dots,n\}
$
to describe an oscillator cluster with $\var_{\alpha}$, $m_{\alpha}$, $c_{\alpha}$, and $k_{\alpha}$ describing the position,  mass, damping, and stiffness of  of the $\alpha$-th scatterer, respectively.
Each oscillator  (scatterer)  possesses a single degree of freedom in the transverse (out-of-plane) direction, $w_{\alpha}(t)$, which describes the displacement of its mass $m_{\alpha}$.
Assuming harmonic motion everywhere in $\Omega$ , we may model the complex deflections of the continuum $\psi(\var)$ as,
\begin{equation}
	\left(\mathcal{D}\nabla^4 - \rho'\omega^2\right)\sol(\var) = f_0\delta({\var-\var_0})+ \sum_{\alpha=1}^n f_{\alpha}\delta(\var-\var_{\alpha}).
	\label{EQ: general_prob}
\end{equation}
where $\mathcal{D} = H^3E/[12(1-\nu^2)]$ is the plate rigidity parameterized by the Poisson ratio $\nu$, thickness $H$, and modulus, $E$, 
and the density per unit area $\rho'$. 
The terms $f_{\alpha}$ describe the complex steady-state forces exerted onto $\psi(\var_\alpha)$ by the $\alpha$-th scatterer. For scatterers modeled as linear spring-mass dampers, $f_{\alpha}$ is given by,
%
\begin{equation}
	f_{\alpha}= k_{\alpha}(w_{\alpha}- \sol(\var_{\alpha}))+i\omega c_{\alpha}\left({w}_{\alpha}- \sol(\var_{\alpha})\right).
	\label{EQ: f_lin} 
\end{equation}
%
Steady-state solutions to~\eqref{EQ: general_prob} take the form,
\begin{equation}
	\sol(\var;\scatterers) = \sol_0(\var) + \psi_{\rm s}(\var;\scatterers),
	\label{EQ: MMS_sol}
\end{equation}
where $\sol_0(\var)$ and $\psi_{\rm s}(\var;\scatterers)$ correspond to the incident solution (which only depends on the incident force) and scattering solutions (which also depends on the set $\scatterers$). 
These solution fields may be expressed in terms of the harmonic Green's function response at each respective point load,
\begin{align}
	\sol_{0}(\var) &= G_{\omega}(\var,\var_0)f_0	\\
	\sol_{\rm s}(\var;\scatterers)&= \sum_{\alpha=1}^{n}f_{\alpha}G_{\omega}(\var,\var_{\alpha}),
	\label{EQ:ScatterForce}
\end{align}
with
the infinite-space Green's function of the Kirchhoff-Love equation given as,
\begin{equation}
	G_{\omega}(\var_{\alpha}, \var_{\beta}) = \frac{i}{8k^2(\omega)}\left(
	H_0(k(\omega)||\var_{\alpha}-\var_{\beta}||)- H_0(ik(\omega)||\var_{\alpha}-\var_{\beta}||)
	\right),
	\label{EQ:Gfun}
\end{equation}
with $H_0(\square)$ being the zeroth-order Hankel function of the first kind and $k(\omega) = [ \rho' H \omega^2/ \mathcal{D}]^{1/4}$ the wavenumber.
To this point, the scattering forces $f_{\alpha}$ of Eq~\eqref{EQ:ScatterForce} are still unknown. However, since the dynamics of the oscillator cluster is both linear and stationary, the forces they exert onto $\sol(\var_{\alpha})$ may be related to the continuum boundary displacement by,
\begin{equation}
	f_{\alpha} =  m_{\alpha}\omega^2\frac{m_{\alpha}\omega_{\alpha}^2 + 	i\omega c_{\alpha} }{m_{\alpha}(\omega_{\alpha}^2-\omega^2  ) + i\omega c_{\alpha}}\sol(\var_{\alpha}) = \mu_{\alpha} \sol(\var_{\alpha}).
	\label{EQ: mu}
\end{equation}
where $\mu_{\alpha}$ is the linear base-motion transmission coefficient.
To solve for the $n$ unknown forcing coefficients $f_1,\ f_2,\dots,\ f_n$, one may construct a system of linear equations by evaluating~\eqref{EQ: MMS_sol} at each oscillator location:
\begin{equation}
	\sol(\var_{\alpha}) = \sol_0(\var_{\alpha}) + \sum_{\beta=1}^{n}f_{\beta}G_{\omega}(\var_{\alpha},\var_{\beta}).
	\label{EQ: Self_cons}
\end{equation}
By substituting $\sol(\var_{\alpha})$ with $\mu_{\alpha}^{-1}f_{\alpha}$, Eq~\eqref{EQ: Self_cons} may be rewritten in compact matrix form as,
\begin{equation}
	\textbf{A}\bm{f} = \textbf{b},
	\label{EQ: Afb}
\end{equation}
where $\textbf{b} = f_0\textbf{G}_{\omega}^0$ is the vector of incident solutions at the oscillator positions, $\textbf{b}_{\alpha}=\psi_0(\var_\alpha)$, and the matrix 
	$\textbf{A} = \hat{\bm{\mu}} - \textbf{G}_{\omega}^{\alpha\beta}$
accounts for the interactions of $\scatterers$ and $\psi(\var)$. The terms $\textbf{G}_{\omega}^0$, $\hat{\bm{\mu}}$, and $\textbf{G}_{\omega}^{\alpha\beta}$  take the definitions,
\[
\textbf{G}^0_{\omega} = 
\begin{bmatrix}
	G_{\omega}(\var_1,\var_0)\\
	G_{\omega}(\var_2,\var_0)\\
	\vdots\\
	G_{\omega}(\var_n,\var_0)\\
\end{bmatrix}, \ 
\hat{\bm{\mu}} =
\begin{bmatrix}
	\mu_1^{-1}&&&\\&\mu_2^{-1}&&\\ &&\ddots&\\&&&\mu_n^{-1}
\end{bmatrix},\  
\textbf{G}_\omega^{\alpha\beta} = \begin{bmatrix}
	G_{\omega}(\textbf{0})&G_{\omega}(\var_1,\var_2)&\dots&G_{\omega}(\var_1,\var_n)\\
	G_{\omega}(\var_2,\var_1)&G(\textbf{0})&\dots&G_{\omega}(\var_2,\var_n)\\
	\vdots	&\vdots&\ddots&\vdots\\
	G_{\omega}(\var_n,\var_1)&\cdots&\cdots&G(\textbf{0})\\
\end{bmatrix},
\]
and may be evaluated directly by use of Eqs~\eqref{EQ:Gfun} and \eqref{EQ: mu}. The implication is that an analytical solution to Eq~\eqref{EQ: general_prob} is provided by solving Eq~\eqref{EQ: Afb} and evaluating Eq~\eqref{EQ: MMS_sol}.

\subsection{Inverse Design Problem Statement}
\begin{figure}[t!]
	\includegraphics[width=\linewidth]{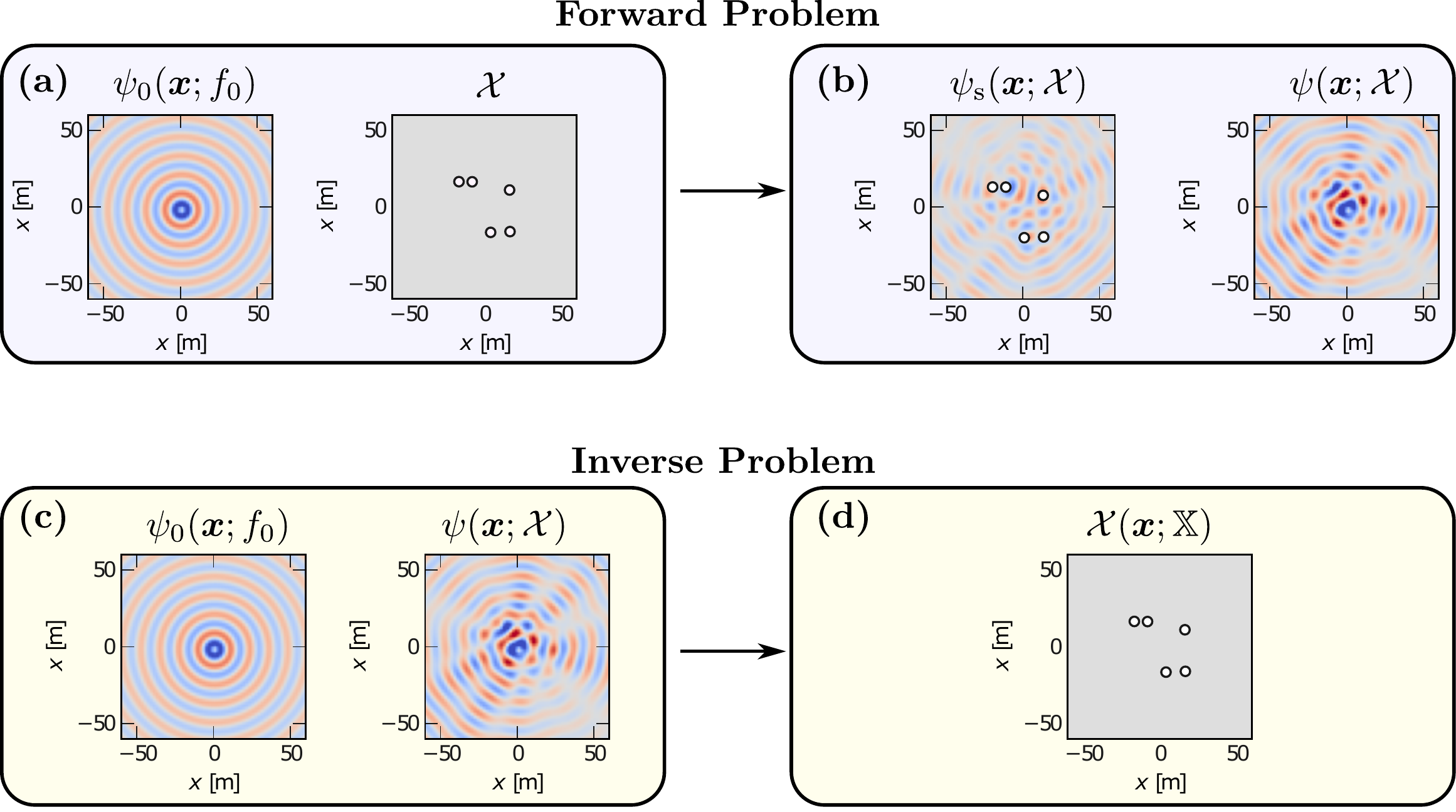}
	\caption{Multiple scattering formulations for forward ((a) and (b)) and inverse ((c) and (d)) formulations. }
	\label{FIG: ProbStatement}
\end{figure}
The analytical formulation of Eq~\eqref{EQ: MMS_sol} addresses the \textit{forward problem} of a multiple scattering field, that is, to resolve $\psi(\bm{x};\scatterers)$ for a given incident force and known scattering cluster (Fig~\ref{FIG: ProbStatement}(a) and (b)).
The \textit{inverse problem} is to reverse this mapping, that is, to construct a function capable of solving for the scattering cluster given that the incident force is known and total wavefield solution viewable in some observable domain $\Omega_{\rm o}$ (Fig~\ref{FIG: ProbStatement}(c) and (d)).
Thus, we pose the modeling goal as follows:
Given an observable field $\Omega_{\rm o}$ that possesses the solution $\psi(\scatterers;\var)$, recover a function that takes the incident wavefield and scattering solution pattern as inputs, and returns an estimate of the scattering array $\mathcal{X}\in\Omega_{\rm s}$ needed to realize $\psi(\var;\scatterers)$ for a given $\psi_0(\var;f_0)$.
Mathematically, we seek the function
\begin{equation}
		\mathcal{M}(\mathbb{X})=\hat{\scatterers},\ \ 
		\mathcal{M}:\Omega_{\rm o}\to\Omega_{\rm s},  \ \  \ \mathbb{X} = \{ {\rm Re}(\psi_0(\bm{x};f_0)), |\psi(\bm{x};\mathcal{X})| \}, 
	\label{EQ: InverseStatement}
\end{equation}
where $\mathbb{X}\in\Omega_{\rm o}$ denotes the observable inputs to a model $\mathcal{M}$, and $\hat{\mathcal{X}}\in\Omega_{\rm s}$ is the predicted scattering set it gets mapped to. We assume the mass, stiffness, and damping of the oscillators to be known, and leave their positions $\var_{\alpha}$ as the unknown quantities of $\scatterers$ that $\mathcal{M}$ must solve for.

Because we are interested in passively managing energy with the multiple scattering clusters, we select the absolute value of the total wavefield $|\psi|=|\psi_{0}+\psi_{\rm s}|$ as the learnable data.
This is also for technical relevance since the amplitude of the total solution field  $\psi$ is of highest importance for wave control applications as compared to  $\psi_{\rm s}$ alone.
Because we desire a model that can generalize across wavenumber, we augment the input data to include the real part of the incident solution (which is assumed known) so that the relationship between incident wavenumber, scattering wavefield energy, and scattering arrangement may be learned by the model.
Hence, by recovering a model described by Eq~\eqref{EQ: InverseStatement}, we aim to accomplish the following modeling goals:
\begin{itemize}
	\item \textit{Scatterer Detection}: To detect the location of scattering elements for a given multiple scattering solution field based on an observed amplitude distribution and known incident wavefield.
	\item \textit{Inverse Design}: To produce a  scattering cluster that is physically consistent with the observed wavefield, that is, a cluster which produces the target field when evaluated by the forward solution.
	\item \textit{Wavefield Engineering}: To produce a scattering cluster that is capable of approximating a desired energy distribution (synthetically generated) for a given incident force.
\end{itemize}

\subsection{Considered Multiple Scattering Problem Types}
\begin{figure}[t!]
	\includegraphics[width=\linewidth]{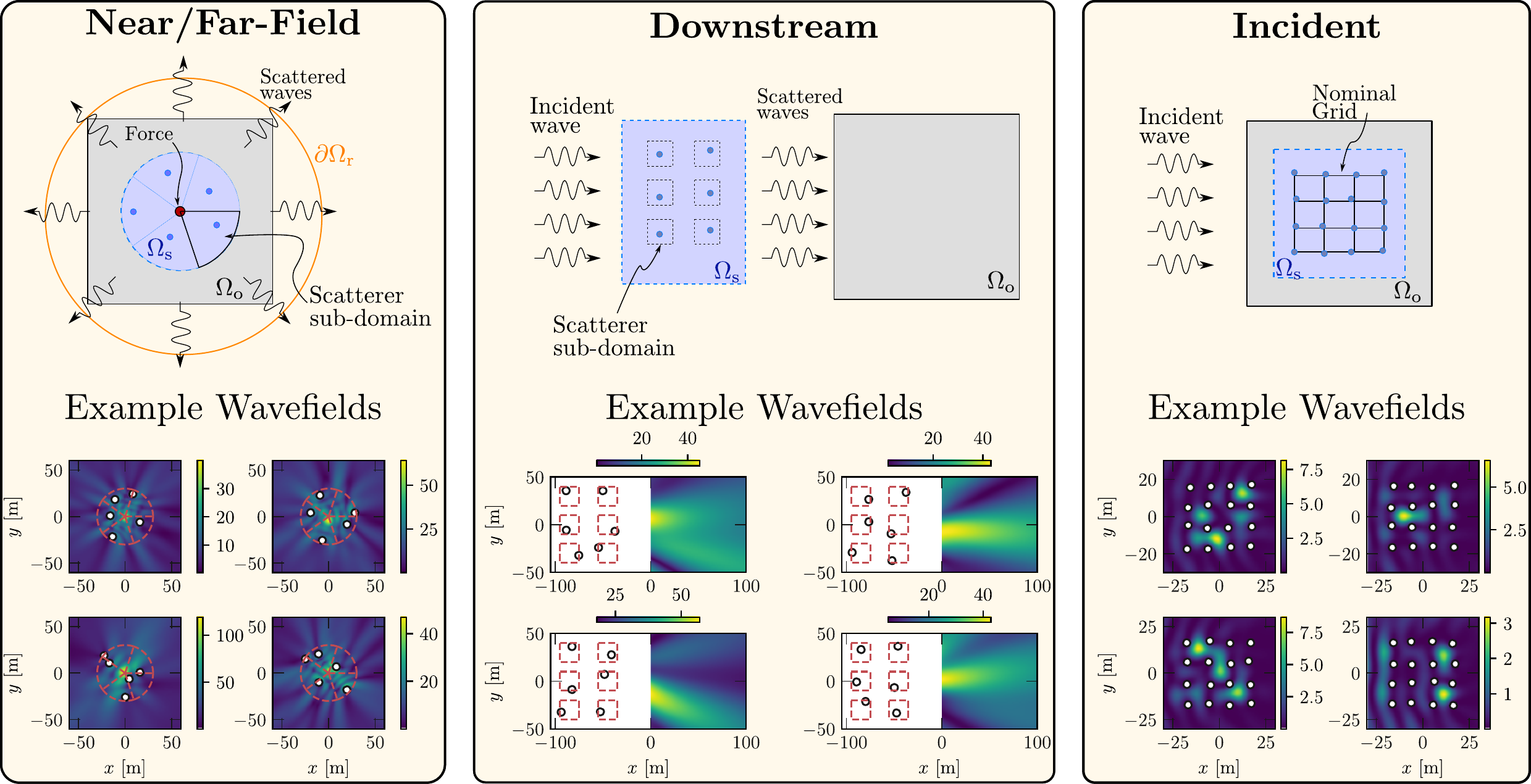}
	\caption{The three problem types considered herein depicting the observable domain $\Omega_{\rm o}$, scattering cluster domain $\Omega_{\rm s}$, and direction of wave propagation; four representative wavefield patterns for each problem type are provided beneath each problem schematic. }
	\label{FIG: ProbTypes}
\end{figure}

We consider the three general classes of multiple scattering problems depicted in Fig~\ref{FIG: ProbTypes}, with each possessing their own intended functionality and corresponding design constraints. 
First, we consider a near/far-field problem whereby incident forces are centered in the observable domain with the scattering array radially enclosing these forces. 
The main purpose of this configuration is to demonstrate the capacity of the proposed model to accurately capture scattering positions in the near-field, and produce oscillator clusters that agree with the ground-truth radial propagation in the far-field.
We next consider a downstream problem whereby incident forces occur in the far field to produce a wave front that passes through a scattering cluster before being observed downstream.
This formulation serves to demonstrate that the proposed model can predict the scattering locations based on downstream wavefield information (e.g., remote sensing) and that the model can design a cluster that steers energy from an incoming incident field to preferred directions.
Finally, we consider an incident localization problem whereby incoming waves interact with an array of scatterers that are perturbed about a nominal grid. This formulation allows for the localization within the grid channels, with the regions of localization being seemingly random functions of the array perturbations as in Anderson localization~\cite{Anderson1958}. 
We therefore seek to inverse design localized wavefields by learning the perturbations that lead to such localization, and then utilize the learned relationships to engineering localization patterns into the wavefield given a predefined incident wave.
Typical wavefields produced by each problem type are give in Fig~\ref{FIG: ProbTypes}, and the details regarding the specific design constraints for each problem type are discussed in the subsequent subsections.


\subsubsection{Near/Far-field Problem Design}
For the near/far-field problem, we consider an incident force of 100 N  applied near the center of the observable domain at $\var_0 = (0,0)+\Delta\var_0$, where $\Delta\var_0$ is drawn from normal distribution with zero mean and variance of 5 meters to introduce a level of randomness in the force location.
Incident wavenumbers are drawn from a random uniform distribution such that  $k\in\left[\pi/10,\,\pi/5\right]$.
An array of five scatterers is considered such that each scatterer is positioned within a sub-quadrant of a circle with radius $r = 30$ meters, making $\Omega_{\rm s}$ a circle centered at the origin. 
The domain $\Omega_{\rm o}$ is centered at the origin and extends 60 meters in each direction, so that both the forcing and scattering domain $\Omega_{\rm s}$ are within $\Omega_{\rm o}$.
This results in wave patterns that propagate radially from the origin which are scattered by the elements of $\scatterers$ before propagating into the far-field.
The goal of this problem design is to serve as a proof-of-concept, e.g., to demonstrate the efficacy of the proposed model to return predicted scattering clusters that have little discrepancy from the target scattering problem both in terms of proximity of scatterer predictions and in terms of the resulting wavefield reconstructed by the predicted scattering arrangement.

\subsubsection{Downstream Problem Design}
For our second problem class we consider an excitation at $\var_0 = (-125,0)$ meters with varying wavenumber $k\in\left[\pi/16,\,\pi/8\right]$.
The incident wave passes through a 6$\times$3 grid of scatterers that is centered about $\var_\mathrm{c}=(-65,0)$ and has a nominal spacing of 40 meters in the $x_1$ direction and 25 meters in the $x_2$ direction. 
We allow for perturbations of the nominal scatterer locations that are evenly distributed within $\Delta \var_\alpha \in \left[-8,\, 8\right]\times\left[-8,\, 8\right]$, resulting in a  56$\times$66 meter scatterer domain $\Omega_{\rm s}$.
In contrast to the Near/Far-field Problem, we now place the 100$\times$100 meter observable domain $\Omega_{\rm o}$ \textit{downstream}, 37 meters to the right side, of $\Omega_\mathrm{s}$.
The perturbations to the nominal scatterer locations result in streams of high energy  passing through $\Omega_\mathrm{o}$ into different directions.
Our goal is to detect resonators that are not within $\Omega_{\rm s}$, and to design resonator arrays that lead to high-energy streams pointing towards given target directions based on observations of downstream scattering solutions.
This includes the goal of downstream wavefield engineering, which can be achieved by evaluating the inverse model on an synthetically engineered energy patterns.
Since $\Omega_{\rm o}$ and $\Omega_{\rm s}$ are disjoint, the model is not able to see the typical patterns/ripples in $\sol$ that form closely around each scatterer.

\subsubsection{Incident Localization Problem Design}
For the downstream problem formulation, we consider farfield forcing at location $\var_0 = (-10^{5},0)$ to produce a near ideal plane wave front with excitation frequency corresponding to an incident wavenumber of $k=\pi/10$.
The scattering cluster is set up around a nominal $4\times4$ grid with uniform 11-meter spacing in the $x$ and $y$ directions and centered at the origin. 
Small variations of $\Delta\var_{\alpha}\in[-1,1]\times[-1,1]$ meter are applied to the nominal coordinate of each scatterer to produce a slightly-perturbed scattering array. 
We consider a 60$\times$60 meter observable domain $\Omega_{\rm o}$ centered at the origin, so that $\Omega_{\rm s}\subset\Omega_{\rm o}$ for the incident problem.
The perturbations to the nominal scattering array result in random localization patterns within the incident wavefield (see Fig~\ref{FIG: ProbTypes}); note that this localization phenomenon is contingent on the incident wavenumber and nominal spacing. Because of this, we do not vary the wavenumber for this problem type.
The goal is to produce a model that is capable of observing localization patterns in the wavefield and predicting the perturbations required to achieve such patterns. Moreover, we seek to engineer customized (synthetic) localization patterns into the observable domain based on the learned relationships between the applied perturbations and the resulting localization patterns. 

\subsubsection{Synthetic Wavefield Design}
\label{SEC:synth_targets}
We now address the training strategy implemented to achieve the final modeling goal of this work, that is,
to \textit{engineer} scattering fields to achieve a desired energy pattern.
Herein, we focus on both directing (downstream problem) and localizing (incident problem) wave energy, as these effects have ubiquitous utility in the context of energy suppression or harvesting~\cite{Wei2017}.
To do this, we must generate synthetic wavefields that are practically realizable, e.g., that are derived to produce patterns similar to wavefield solutions. 
To this end, we implement two approaches for synthetic wavefield designs, depending on the design goal.

\begin{figure}[t!]
	\begin{subfigure}{.5\linewidth}
		\includegraphics[width=\linewidth]{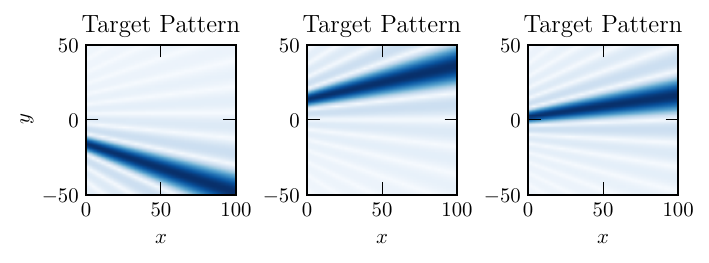}
		\caption{Downstream target patterns}
	\end{subfigure}
	\begin{subfigure}{.5\linewidth}
		\includegraphics[width=\linewidth]{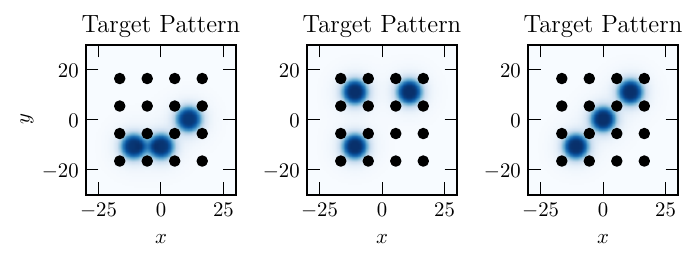}
		\caption{Incident target patterns}
	\end{subfigure}
\caption{Example of downstream target patterns for (a) downstream wave steering as generated per Eq~\eqref{EQ: Synth_DS} and (b) incident wave localization as generated by Eq~\eqref{EQ: Synth_Inc} with the nominal incident oscillator grid superimposed.
}
\label{FIG:Synth_Targs}
\end{figure}

For the downstream problem, we seek to direct energy streams along a target angle $\varphi$ from a given center location $\var_{\rm sc}$. 
Similar to the design approach in \cite{Packo2021}, we achieve this by using a Fourier expansion with of an angular Dirac impulse $|\sol(\var)|=\delta(\mathrm{arg} \left\{\var-\var_{\rm sc}\right\}-\varphi)$, where $\mathrm{arg} \left\{\Box\right\}$ denotes the angle defined between the input and the positive $x_1$-direction. The series expansion up to the harmonic order $H=40$ then reads,
\begin{equation}
	|\overline{\sol}(\var)| =  \left| \sum_{h=0}^H \ee^{i \cdot h \cdot (\mathrm{arg}\left\{ \var -\var_{\rm sc} \right\} -\varphi)} \right| ,
	\label{EQ: Synth_DS}
\end{equation}
where we use $\overline{\psi}$ to denote the synthetic field. We generate different realizations by randomly varying $\varphi \in [-\pi/8,\pi/8]$ and centering at $\var_{\rm sc} =(-60,0)$. 
For the incident problem, we are interested in scatterer arrays that lead to a localization of energy within predetermined channels of the nominal cluster. Similar synthetic fields that are localized in $N$ channels can be approximated as,
\begin{equation}
	|\overline{\sol}(\var)| =  \sum_{j=1}^N 1- \frac{1}{1+\left(\frac{R}{||\var-\var_{{\rm sc},j}||}\right)^M} \, ,
	\label{EQ: Synth_Inc}
\end{equation}
where $\var_{{\rm sc},j}$ is the center coordinate of the $j$-th channel. Moreover, the localization radius $R$ determines how wide the localized area in the channel is, while the exponent $M$ governs how quickly the energy decays around the localized area. For each generated synthetic field, we randomly select $N\in\left[1,\,9\right]$ channels with an exponent of $M=4$ and a radius of $R=5$.
Examples of synthetic fields generated by Eqs~\eqref{EQ: Synth_DS} and~\eqref{EQ: Synth_Inc} are depicted in Fig~\ref{FIG:Synth_Targs}(a) and~\ref{FIG:Synth_Targs}(b), respectively.

\section{Machine Learning Framework}
\label{Sec:S3_ML}

The objective of the machine learning model is to utilize information from the observed wavefield solutions $\sol(\var;\scatterers)$ (or synthetic targets $\overline{\sol}(\var)$) to predict the scattering coordinates $\hat{\scatterers}$ believed to have produced the wave patterns.
To this end, the model must first featurize the wavefield information, that is, extract information from the observable scattering field, and then to map it to the appropriate set of scattering coordinates.
In the following discussion, we detail the architecture, loss functions, and training methodology utilized to optimally construct this mapping in the context of multiple scattering inverse design.

\subsection{Architecture}
\begin{figure}[t!]
	\includegraphics[width=\linewidth]{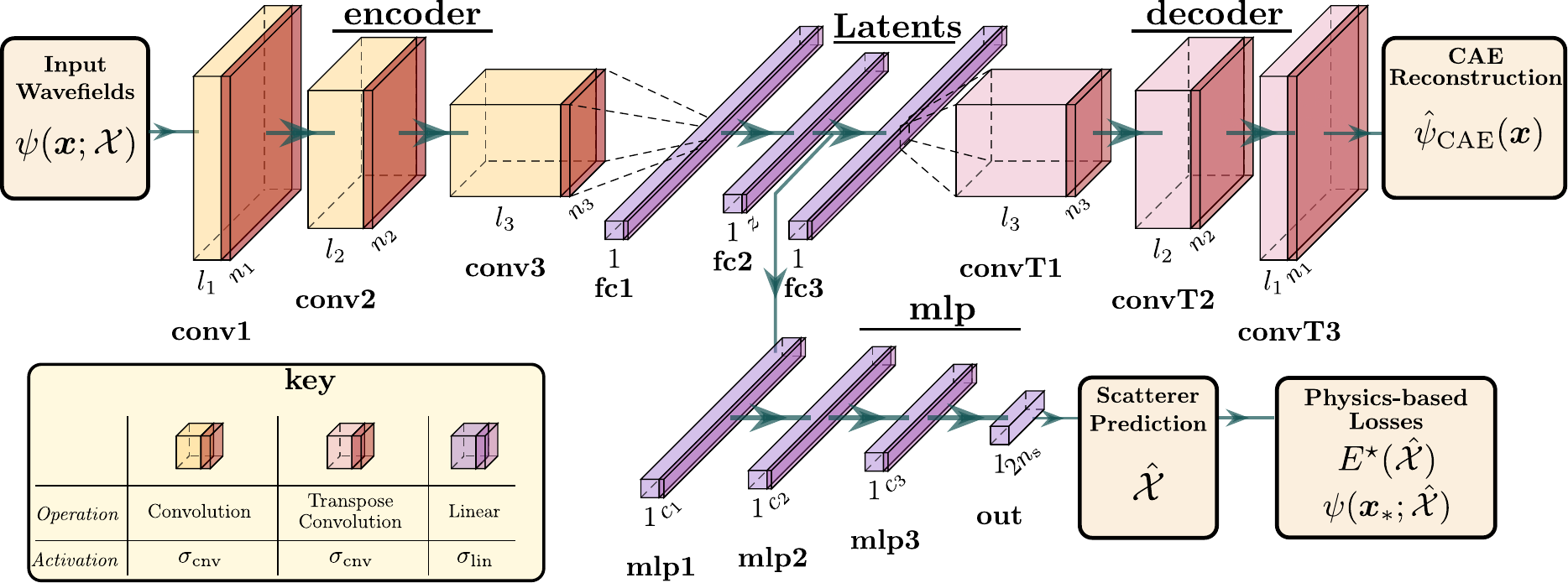}
	\caption{Schematic of the proposed CAE-MLP workflow.  The encoder of the CAE featurizes and compresses scattering wavefields into the latent vector $\bm{z}$, the MLP predicts a scattering cluster $\hat{\scatterers}$ based on  $\bm{z}$, and the physics-informed loss quantities $E^\star(\hat{\scatterers})$ and $\psi(\var_*;\hat{\scatterers})$ are subsequently computed.}
	\label{FIG:  NN_Schematic}
\end{figure}
We adopt a deep convolution auto-encoder (CAE) model to generate a low dimensional latent representation of the input wavefields, and we coupled this to a physics-informed multi-layer-perception (MLP) module that is responsible for predicting the scatterer locations; a schematic of the model architecture is depicted in Fig~\ref{FIG: NN_Schematic}.
The CAE consists of an encoder, latent dimension, and decoder; it's primary purpose is to produce informative latent dimensions for the MLP to use. The MLP consists of a series of fully connected layers which are responsible for mapping the latent dimensions to a scattering cluster prediction. 
We note that the activation functions for the convolutional layers ($\sigma_{\rm cnv}$) may differ from those selected for fully connected linear layers ($\sigma_{\rm lin}$).

The encoder featurizes the wavefield with forward convolutions (denoted as ``conv" in Fig~\ref{FIG: NN_Schematic}) and \textit{compresses} the feature maps to a latent vector $\bm{z}$ with fully connected layers (denoted as ``fc" in Fig~\ref{FIG: NN_Schematic}). 
The decoder is tasked with the inverse operation of approximating the input tensors;
it does so by lifting the dimension of $\bm{z}$ with fully connected layers based on the final convolutional feature maps, and applying transpose convolutions (denoted ``convT" in Fig~\ref{FIG: NN_Schematic}) until the output matches the dimension of the input.
The decoder output is not directly utilized in the objective of scatterer prediction it is a pure data mapping. However, jointly optimizing the decoder with the rest of the network encourages a model parameterization that produces informative latent vectors in the context of wavefield features.

The MLP module is responsible for the physically relevant predictions in the context of remote sensing and inverse design.
It utilizes a series of fully connected layers to map the low-dimensional latent representation of the wavefield to a set of predicted scattering coordinates, $\hat{\scatterers}$.
Altogether, the mapping from the input of the CAE to the output of the MLP delivers the desired inverse function approximation of Eq~\eqref{EQ: InverseStatement}.
Since an analytical forward model is known for the scattering set, we may construct physically meaningful quantities from the MLP output and use them to inform the training of the MLP and encoder; this provides the relevant physical constraints to the network to make it \textit{physics-informed}.
These physics-based loss quantities which are detailed in the subsequent section.


\subsection{Loss Functions}
\label{Subsec: losses}
For the training of our network, we consider a  joint loss function $L(\paramvec;\mathbb{X},\hat{\mathbb{X}},\scatterers,\hat{\scatterers})$  composed of a weighted sum of both data-based and physics-based penalty terms:
\begin{equation}
	L(\paramvec;\mathbb{X},\hat{\mathbb{X}},\scatterers,\hat{\scatterers}) = \lambda_1 L_\mse(\paramvec;\mathbb{X},\hat{\mathbb{X}})
	+   \lambda_2 L_\mse(\paramvec;\scatterers,\hat{\scatterers}) 
	+	
	\lambda_3L_{\rm force}(\paramvec;\scatterers,\hat{\scatterers})  
	+ 
	\lambda_4L_{\rm sparse}(\paramvec;\scatterers,\hat{\scatterers}) \, .
	\label{EQ: JointLoss}
\end{equation}
The first data loss, $L_\mse(\paramvec;\mathbb{X},\hat{\mathbb{X}})$, with weight $\lambda_1$ is the commonly-used Mean-Squared-Error (MSE) loss between the input tensors $\mathbb{X}$ and the decoder output $\hat{\mathbb{X}}$,
\begin{equation}
	L_\mse(\paramvec;\mathbb{X},\hat{\mathbb{X}}) =   \frac{1}{n}\sum_{k=1}^n {||\mathbb{X}_k - \hat{\mathbb{X}}_k||^2} \, .
	\label{EQ: L2Wave}
\end{equation}
Herein, $||{\Box}||$ refers to the $L_2$ vector norm of $\Box$.
The remaining losses all pertain to the scatterer predictions of the MLP module. We apply the same MSE loss quantity with corresponding weight $\lambda_2$ in order to administer a data-driven penalty during training: 
\begin{equation}
	L_\mse (\paramvec;\scatterers,\hat{\scatterers}) =  \frac{1}{n}\sum_{k=1}^n {||\scatterers_k - \hat{\scatterers}_k||^2} \, .
	\label{EQ: L2Res}
\end{equation}

The analytical structure of the multiple scattering forward solutions grants us unique access to physics-based penalty terms. To this end, we formulate two physics-based penalty functions, both being derived from multiple scattering theory. 
We term first physics-based loss as the forcevector loss, $L_{\rm force}(\paramvec;\scatterers,\hat{\scatterers})$, which we formulate as:
\begin{equation}
	L_{\rm force}(\paramvec;\scatterers,\hat{\scatterers}) =  \frac{1}{n{f_0^2}}\sum_{k=1}^n {|E^\star(\scatterers_k) - E^{\star}(\hat{\scatterers}_k)|} + \underbrace{\frac{1}{n}\sum_{k=1}^n  \frac{\left\Vert \mean{\alpha}{\var_{\alpha}\in \scatterers_k} - \mean{\alpha}{ \hat{\var}_\alpha \in \hat{\scatterers}_k }  \right\Vert}{\left\Vert \mean{\alpha}{\var_{\alpha}\in \scatterers_k} - \var_0  \right\Vert}}_\text{centroid regularization} \, .
	\label{EQ: ForeceVectorLoss}
\end{equation}
This is defined as the absolute deviation between the energy quantities $E^\star$ which can be computed from the solution of Eq~\eqref{EQ: Afb} as:
\begin{equation}
	E^\star(\scatterers) =  \textbf{b}^{\dagger}(\scatterers)\textbf{A}^{-1}(\scatterers)\textbf{b}(\scatterers)= \textbf{b}^{\dagger}(\scatterers)\textbf{f}(\scatterers) \, .
	\label{EQ: ForeceVectorLoss_AFb}
\end{equation}
This scalar quantity encodes the interactions between all point loads in $\Omega$ (those being the incident force and scattering forces). 
Since it is only a function of the relative distances between the scatterers and their distances to the forcing, it is invariant to rotations of the entire array of scatterers around the forcing location.
We therefore introduce an additional regularization term to $L_{\rm force}$ that returns the absolute error between the centroids of $\scatterers$ and $\hat{\scatterers}$. As rotations around $\var_0$ become more relevant the closer the centroid of the array lies to the forcing location, the error is normalized by the distance of the true array's centroid to $\var_0$.
Only in the special case where the centroid of $\scatterers$ coincides with the forcing location is the regularization term is singular. This does not pose a limitation for the problem classes that we consider in this work, however, we note that this introduces a restriction to generality. 
To keep the deviation between the energy quantities in a similar order of magnitude as the centroid regularization, we obtained good results when normalizing the energy quantities by the squared forcing amplitude $f_0^{2}$.

Our second loss quantity is a direct embedding of the multiple scattering solution into the network by computing the forward solution of the predicted scattering cluster and comparing it directly to the target wavefield.
We term this loss quantity as the sparse-reconstruction loss,
 $L_{\rm sparse}(\paramvec;\scatterers,\hat{\scatterers})$, which is weighted by $\lambda_4$ and computed as:
\begin{equation}
	L_{\rm sparse}(\paramvec;\scatterers,\hat{\scatterers}) = \frac{1}{n}\sum_{k=1}^n || |\sol(\var_*;\scatterers_k)| - |\sol(\var_*;\hat{\scatterers}_k)|||^2 \, .
	\label{EQ: SparseLoss}
\end{equation}
This loss value governs the MSE between the absolute value of the target field $\sol(\scatterers_k;\var_*)$, which can either be a synthetic field or the solution to a true scatterer array, and the absolute value of the wavefield generated with the predicted array $\sol(\hat{\scatterers}_k;\var_*)$.
Since solving the multiple scattering problem over the full pixel grid in $\Omega_{\rm o}$ used to generate $\mathbb{X}$ would be too computationally expensive for an efficient training, we only sample it on a sparse subset of the pixel grid that we refer to as $\var_*$. 
For each batch of data we randomly select a subset of  225 points from the large grid to compute the solution over. 
It should be noted that choosing more or less points in $\var_*$ results in a smoother or rougher loss surface, respectively and thus has a direct influence on the training. 
This loss function is especially helpful when the network is trained on synthetic target fields, as it does not require information about the true scatterer array $\scatterers$. 
However, due to the non-uniqueness of multiple scattering problems, Eq~\eqref{EQ: SparseLoss} may be divergent if training is performed outside a zone of convergence, and therefore it should not be employed until the model has been trained for several epochs with data-driven losses.

The  gradients of Eqs~\eqref{EQ: ForeceVectorLoss} and~\eqref{EQ: SparseLoss} needed to perform backpropagation are listed in \ref{APX: Loss}.
The described architecture and custom loss functions were implemented in PyTorch 2.0.1; we note that the Hankel functions required to evaluate 
Eqs~\eqref{EQ: ForeceVectorLoss} and~\eqref{EQ: SparseLoss} are not available with GPU support in PyTorch.
Therefore, spline-approximations of the Hankel functions were utilized when evaluating the forward and backward pass of the physics-based losses. 
Details on these spline approximation and the confirmation of their convergence are given in~\ref{APX: Spline}.

\subsection{Model Training and Optimization}

With the model architectures and loss functions defined, we now give focus to the training strategy and model paremterization.
Both of these aspects may drastically influence the end performance of the model, and hence we devote explicit attention to developing an appropriate training scheme and optimizing the model paremterization within the context of our aforementioned design goals.
Accordingly, this section outlines the training strategies developed to optimize the model for remote sensing, inverse design, and wavefield engineering, as well as to overview a multi-staged hyperparameter selection scheme developed to parameterize the model.

\subsubsection{Training Methodology}

We adopt a two-stage model optimization routine that is designed to focus first on remote sensing optimization and second on inverse design.
In the first stage, we focus on the $L_{\rm MSE}$ losses and $L_{\rm force}$ to give primary attention to bringing the predicted scattering cluster as close as possible to the true scattering cluster.
Since $L_{\rm sparse}$ may be divergent outside a zone of convergence, this first stage also serves to draw scatterer predictions sufficiently close for Eq~\eqref{EQ: SparseLoss} to be useful.
Moreover, the computation of Eq~\eqref{EQ: SparseLoss} and its gradients are the greatest computational expense in training; hence, our approach provides a relatively inexpensive burn-in phase of training. 
After stage I, the sparse-reconstruction loss is activated to mark the start of stage II; this training stage optimizes the model for the inverse design problem. Due to the non-uniqueness of solutions, this objective is not always aligned with stage I, and therefore the model may produce substantially different predictions depending on whether it is evaluated at the end of stage I or II.
The summary of the two stages are as follows:
\begin{itemize}
	\item \textit{Training Stage I} -- 
	In the fist stage, $L_{\rm sparse}$ is ignored and backpropagation is only contingent on the data-losses of Eqs~\eqref{EQ: L2Wave},~\eqref{EQ: L2Res}, and~\eqref{EQ: ForeceVectorLoss}.
	All parameters are learnable and are updated by 60 epochs considering a weighted combination of the the joint data loss. 
	In principal, this delivers model weights which bring the predicted scattering cluster to be reasonably close to the true scattering array. 
	\item \textit{Training  Stage II} -- After the burn-in stage, $L_{\rm force}$ is activated in Eq~\eqref{EQ: JointLoss}. 
	However, the decoder outputs and gradient contributions are ignored, and Eq~\eqref{EQ: L2Wave} is dropped from the joint loss.
	This stage is run for 40 epochs, which was found to be sufficient for validation loss convergence.
\end{itemize}

During training, the input channels were normalized to posses zero mean and unit variance (z-standard normalization) in order avoid bias. The weights of each layer were initialized using a random uniform distribution centered at zero and with variance inversely proportional to the number of trainable parameters in the layer.
Parameter updates were performed using the ADAM optimizer with learning rates that were determined by the hyperparameter selection scheme. 
Fore each problem type of Fig~\ref{FIG: ProbTypes}, 100,000 data training samples were generated based on a Latin Hyper-Cube sampling of the scattering design space (with an additional 1,000 samples being generated for post-training testing).
Training utilized an 80-20 split of the training and testing data.
The training was implemented on NVIDIA V100 GPUs accessible via the Hardware Accelerated Computing cluster of the National Center for Supercomputing Applications at the University of Illinois Urbana-Champaign~\cite{Kindratenko2020}.

\subsubsection{Transfer Learning for Synthetic Wavefields}
\label{subsec: TFL}
Models that are trained extensively on physical multiple scattering solutions ($\psi(\var;\scatterers)$) would not be expected to perform equally well when evaluating the synthetic fields $\overline{\psi}(\var)$.
This is because the wavefield featurizations that the model learns during training are inevitably linked to nuanced wavefield phenomena that exist in true scattering solutions but not in synthetic target fields.
Therefore, to prime the models to predict effectively on the synthetic fields, we employ the strategy of transfer learning. 
In brief, this methodology is used to re-tune a trained model to predict on data that is outside of the original training distribution~\cite{Weiss2016}.
The main advantage is that we can leverage the wavefield recognition capacity  learned by the models during training phases I and II to rapidly adapt to the new (out-of-distribution) synthetic wavefields without the need for exhaustive retraining.
In this transfer training stage, we utilize only $L(\var) = L_{\rm sparse}(\paramvec;\overline{\psi}(\var_*),\psi(\var_*;\hat{\scatterers}))$ since none of the other loss functions are applicable for synthetic fields. 
The synthetic wavefields were z-standard normalized based on the  normalization previously described so that the input channels do not differ in amplitude distribution compared to the prior training stages.
However, before begin evaluated by the physics-based loss function $L_{\rm sparse}(\overline{\psi}(\var_*),\psi(\var_*;\hat{\scatterers}))$, they are
un-normalized using the mean and standard deviation of the training dataset. This ensures that $|\sol(\hat{\scatterers}_k)|$ used in $L_{\rm sparse}$ is of similar order of magnitude, since the target of transfer learning is to bring $\psi(\var_*;\hat{\scatterers})$ as close as possible to $\overline{\psi}(\var_*)$.
During this stage, we utilize 1,500 synthetic training fields and employ backpropagation over 200 epochs at a learning rate of $\gamma = 2\times10^{-4}$.

\subsubsection{Model Parameterization}
\label{subsec:HyperOpt}
To determine a suitable set of hyperparameters (e.g., learning rates, loss weights, etc.), we adopted a multi-stage hyperparameter optimization routine that was tailored for the specified modeling goals. 
Between the model architecture and training scheme, a total of 16 hyperparameters were selected to be identified through the optimization scheme.
Since hyperparameter optimization is a notoriously costly task, we utilized a Bayesian optimization scheme to optimally sample the hyperparameter space. 
However, the high-dimensionality of the hyperparameter space rendered direct implementation of Bayesian optimization impractical, since too many samples would be needed in order to have confidence in the returned posteriors. 
Therefore, we developed a multi-stage routine whereby each stage is responsible for optimizing a successive subset of the hyperparameters.
We begin by optimizing the CAE kernel widths, channel depths, and activation functions to produce the most informative latent spaces based on the evaluation of the decoder output.
We next optimize the MLP-related hyperparameters such as latent dimension, number of fully connected layers, and MLP activation functions based on the evaluation of the MSE loss of the scatterer predictions (with the physics-losses still being ignored at this stage).
Finally, we optimize the hyperparameters of the physics-based training such as the learning rates,  loss weights, and the batch size, based on a joint-measure considering the scatterer placement accuracy, $\hat{\scatterers}$, and the inverse solution accuracy, $\psi(\var;\hat{\scatterers})$.
Details and results regarding this multi-stage hyperparameter optimization routine are given  in~\ref{APX: Hyperopt}.
	
\section{Results}
\label{Sec:S4_Resuls}

This section presents the results of the modeling methodology of Section~\ref{Sec:S3_ML} applied to the problems depicted by Fig~\ref{FIG: ProbTypes}.
Herein, we utilize the terminology ``target" wavefield to correspond either to a ground truth multiple scattering field ($\sol(\var;\scatterers)$) or synthetic target ($\overline{\sol}(\var)$) that the model assesses.
Reference to predicted scattering clusters denotes MLP predictions, $\hat{\scatterers}$, and reconstructed wavefield refers to the physical evaluation of the predicted cluster, e.g, $\psi(\var;\hat{\scatterers})$. 
This fundamentally differs from the wavefield generated via the decoder output, which we denote as $\hat{\psi}_{\rm CAE}$, since the later is constructed purely from mapping data, while the former is based on the forward physics model applied to the predicted parameters. 

We present both quantitative and qualitative assessments with respect to remote sensing and inverse design. 
We emphasize that while these two goals are seemingly harmonious, they are not the same; remote detection aims to minimize the error between the predicted  and true scattering sets, whereas inverse design aims to minimize the error between the target  and reconstructed wavefields. 
Lastly, we present the performance of the modeling methodology in assessing the synthetic targets described in Section~\ref{SEC:synth_targets} in order to
demonstrate its efficacy  as a design tool for wavefield engineering.

\subsection{Training Summary}
\label{subsec: Training}

\begin{figure}[t!]\centering
	\begin{subfigure}{\linewidth}\centering
		\includegraphics[width=\linewidth]{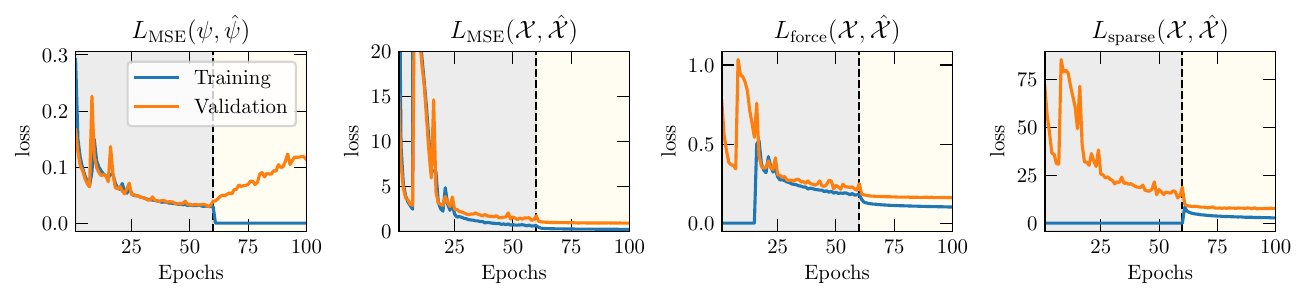}
		\caption{Near/far-field problem}
	\end{subfigure}
	\begin{subfigure}{\linewidth}\centering
		\includegraphics[width=\linewidth]{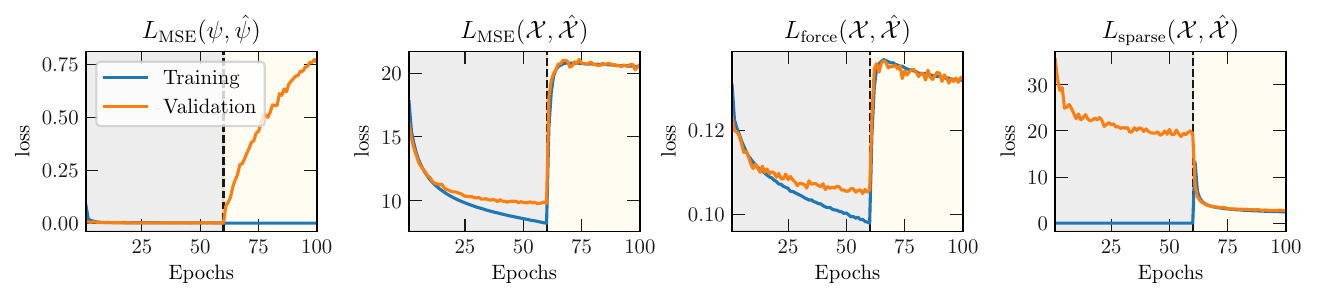}
		\caption{Downstream problem}
	\end{subfigure}
	\begin{subfigure}{\linewidth}\centering
		\includegraphics[width=\linewidth]{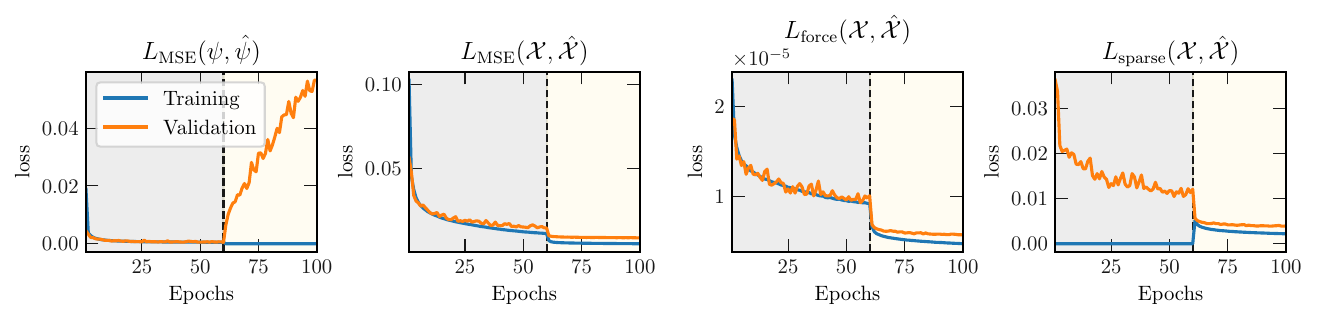}
		\caption{Incident localization problem}
	\end{subfigure}
	\caption{The four loss quantities computed for training and validation data across 100,000 data samples for each problem type for (a) the near/far-field problem, (b) the downstream problem, and (c) the incident localization problem. From left-to-right, the MSE loss of the CAE reconstruction, MSE loss of the MLP predictions, force-vector physics-based loss, and sparse-reconstruction physics-based loss is depicted. The vertical line at epoch 60 indicates the activation of physics loss modules; the two-color indicates this transition as well.}
	\label{FIG: Losses}
\end{figure}

Figure~\ref{FIG: Losses} presents the training histories for the three respective problem types. 
Commenting first on the near/far-field problem, we note that all losses decline during stage I of training (denoted by the shaded gray region). 
Upon activation of the physics-losses at the onset of stage II, a clear dip is observed in the sparse-reconstruction loss profile which converges near the training loss.
In contrast, the MSE loss of the CAE jumps at this point (as expected) since the decoder is not trained during stage II. This phenomenon occurs for each problem considered. 
However, we emphasize that $\L_\mse(\mathbb{X},\hat{\mathbb{X}})$ is of little practical interest since it represents only data-driven reconstruction of the known input field.

Turning next to the downstream problem of Fig~\ref{FIG: Losses}(b), we note similar learning behavior during stage I with the MSE loss of $\scatterers$ declining steadily. 
However, at the onset of stage II, this same loss propels upward, as does the force vector loss; we comment that these two losses are correlated since the measures $\textbf{A}$ and $\textbf{b}$ of Eq~\eqref{EQ: ForeceVectorLoss_AFb} encode the placement of point loads.
Interestingly, as the MSE loss of $\scatterers$ and  the force vector losses increase during stage II, the sparse-reconstruction loss $L_{\rm sparse}(\scatterers,\hat{\scatterers})$ falls significantly.
Finally commenting on the incident problem of Fig~\ref{FIG: Losses}(c), we see similar trends for stage I yet again, that is, consistent validation loss reduction across all losses.
Moreover, the validation loss of the physics-based loss functions drops drastically at the start of stage II, corresponding to their activation in training. 
We comment that validation and training losses remain relatively close to each other for all loss quantities, indicating that over-fitting has been largely avoided.

The above observations offer several notable insights towards the model's functionality at each training stage. 
Stage I is clearly effective in driving down the MSE loss of the scatterers, whereas stage II either offers little improvement to this loss metric, or even degrades it entirely.  
Since multiple-scattering solutions are not unique, it follows that many  scattering configurations may lead to the locally \textit{optimal} inverse designs in the wavefield reconstruction sense, and therefore the local minima found via $L_{\rm MSE}(\scatterers,\hat{\scatterers})$ may conflict with those found by $L_{\rm sparse}(\scatterers,\hat{\scatterers})$. 
Hence, once stage II training commences and the sparse-reconstruction loss contributes to model updating, we see its corresponding validation loss  plummet severely, indicating that the produced cluster designs emulate the desired solution field more faithfully. 
This effect is most profound for the downstream and incident problem types which is largely due to the fact that $\lambda_4$ was selected to be much higher than $\lambda_2$ for these problem types (\ref{APX: Hyperopt}); the implication is that $L_{\rm sparse}$ dominates the learning once activated. 
From the standpoint of remote sensing, $L_{\rm MSE}(\scatterers,\hat{\scatterers})$  is the most important, whereas $L_{\rm sparse}(\scatterers,\hat{\scatterers})$ is the most important for inverse design; the former guides stage I learning while the latter  dominates stage II.
It follows that stage I primes the model for performance in remote sensing, however its inverse design capacity is not optimized until stage II training has completed.
This is behavior is also encoded into the hyperparameter optimization which values the remote sensing capacity at the end of stage I, and the inverse design capacity at the end of stage II (\ref{APX: Hyperopt}).
In summary, we find that the model changes its effective functionality depending on whether it is evaluated at the end of stage I or stage II; we will further demonstrate this in subsequent sections.

\subsection{Scatterer Detection and Inverse Solutions: Quantitative Evaluation}
\label{subsec: Scatterers}
\begin{figure}[t!]
	\centering
	\begin{subfigure}{\linewidth}
		\begin{subfigure}{.5\linewidth}\centering
			\textbf{Stage I Training}\\
			\includegraphics[width=.5\linewidth]{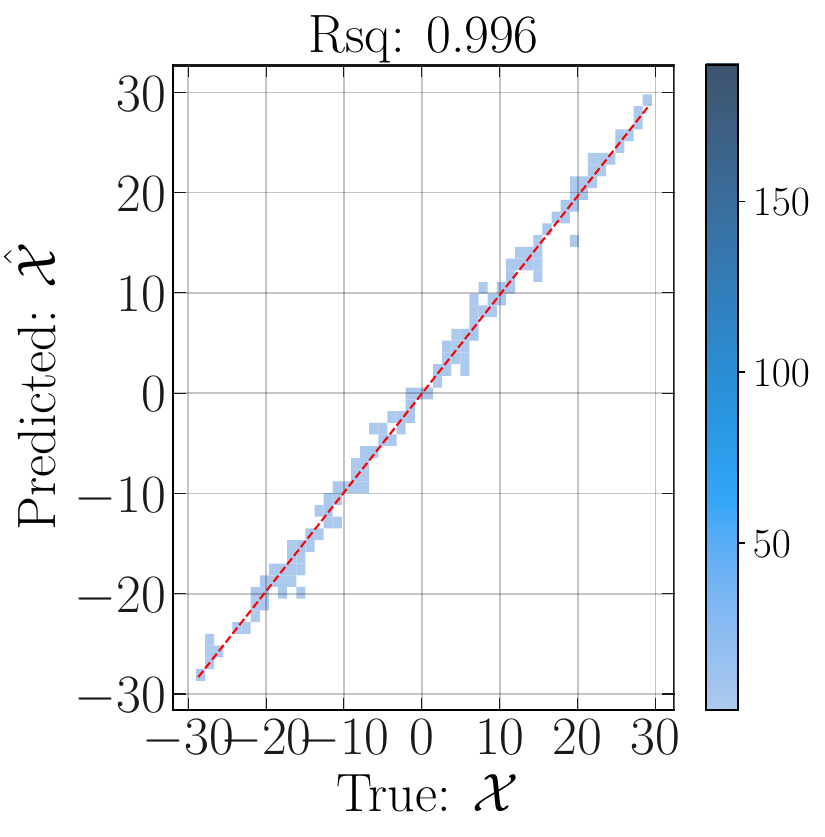}%
			\includegraphics[width=.5\linewidth]{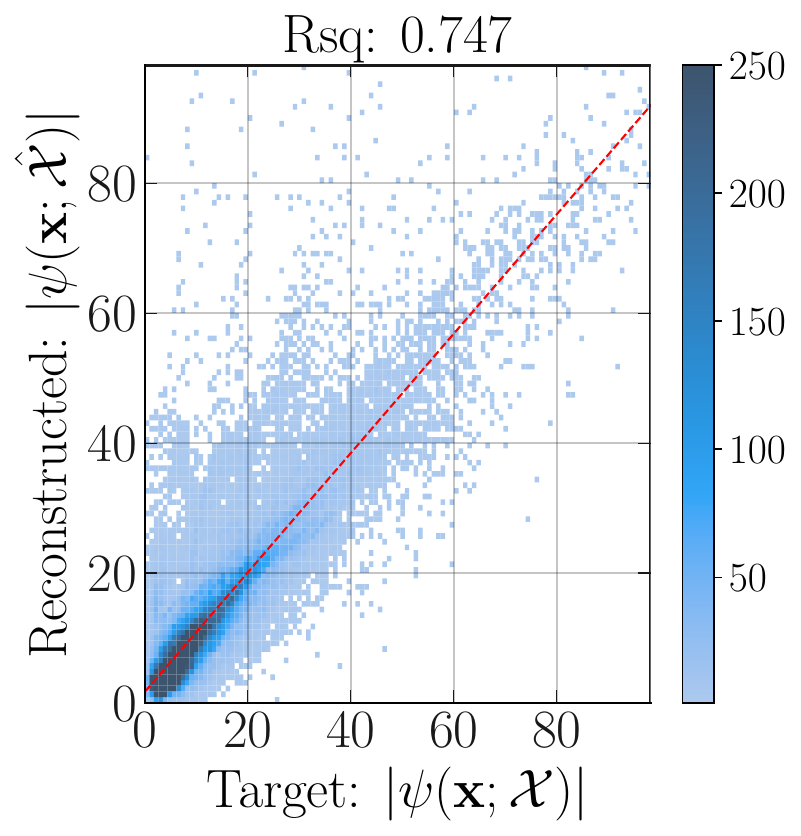}
		\end{subfigure}%
		\begin{subfigure}{.5\linewidth}\centering
			\textbf{Stage II Training}\\
			\includegraphics[width=.5\linewidth]{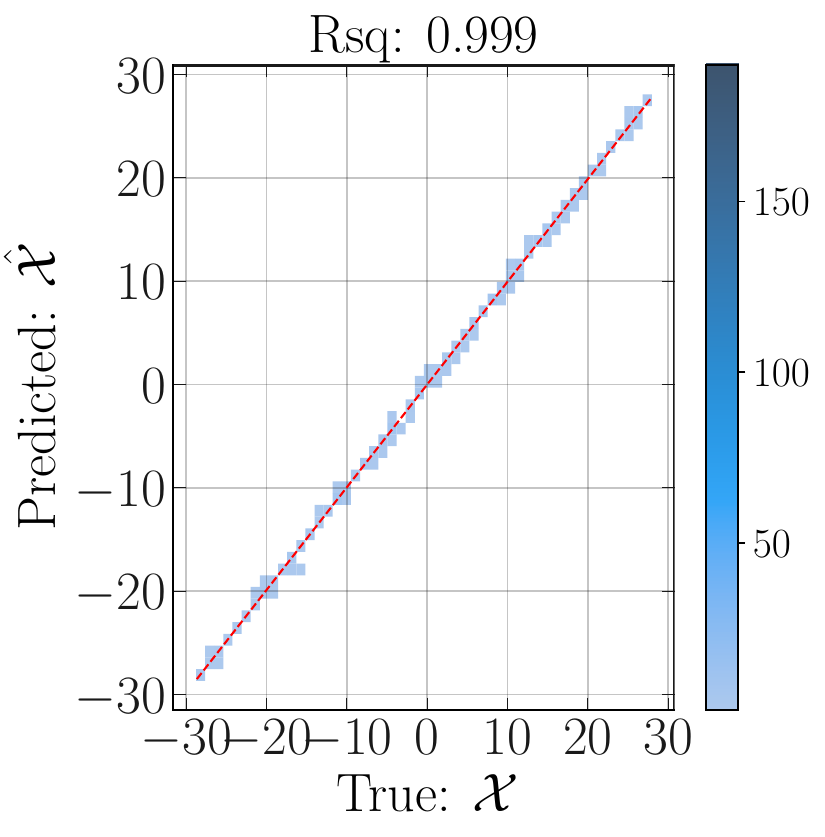}%
			\includegraphics[width=.5\linewidth]{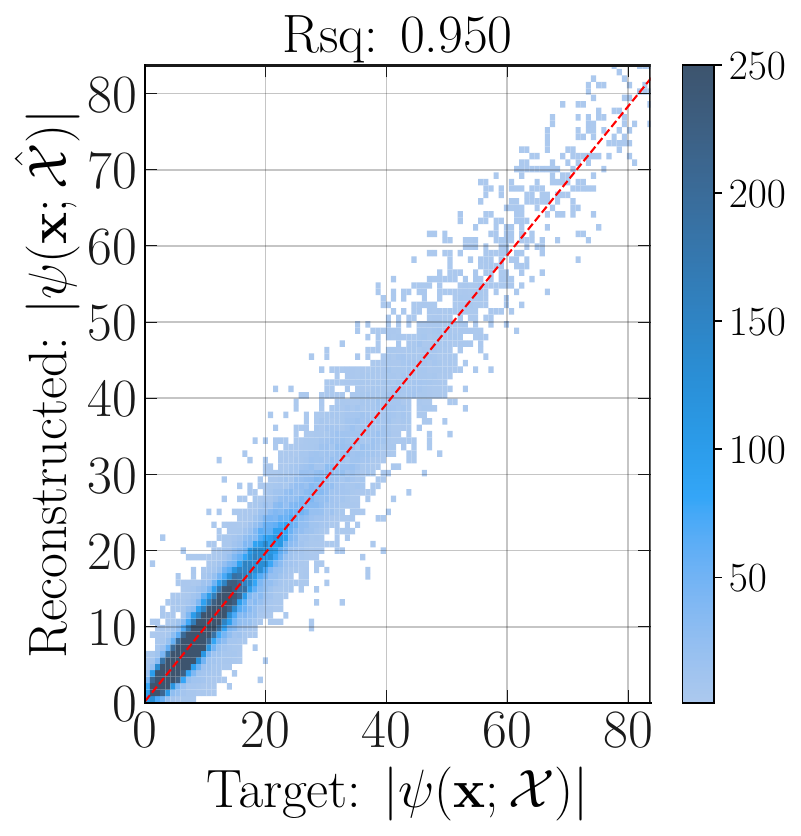}
		\end{subfigure}
		\caption{Near/far-field problem}
	\end{subfigure}
	
	\begin{subfigure}{\linewidth}
		\begin{subfigure}{.5\linewidth}
			\includegraphics[width=.5\linewidth]{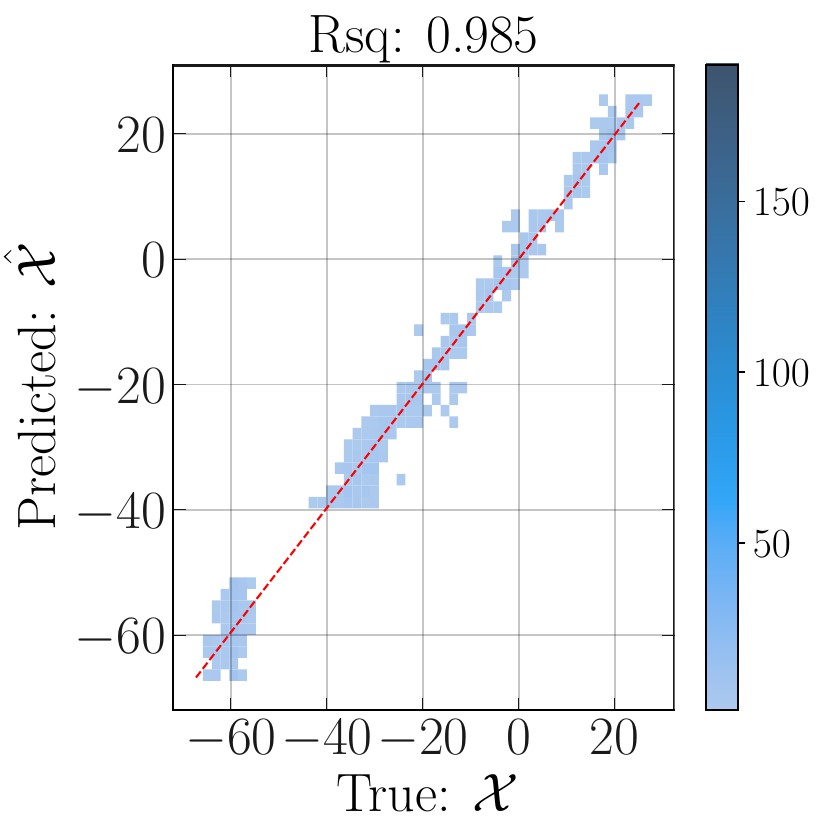}%
			\includegraphics[width=.5\linewidth]{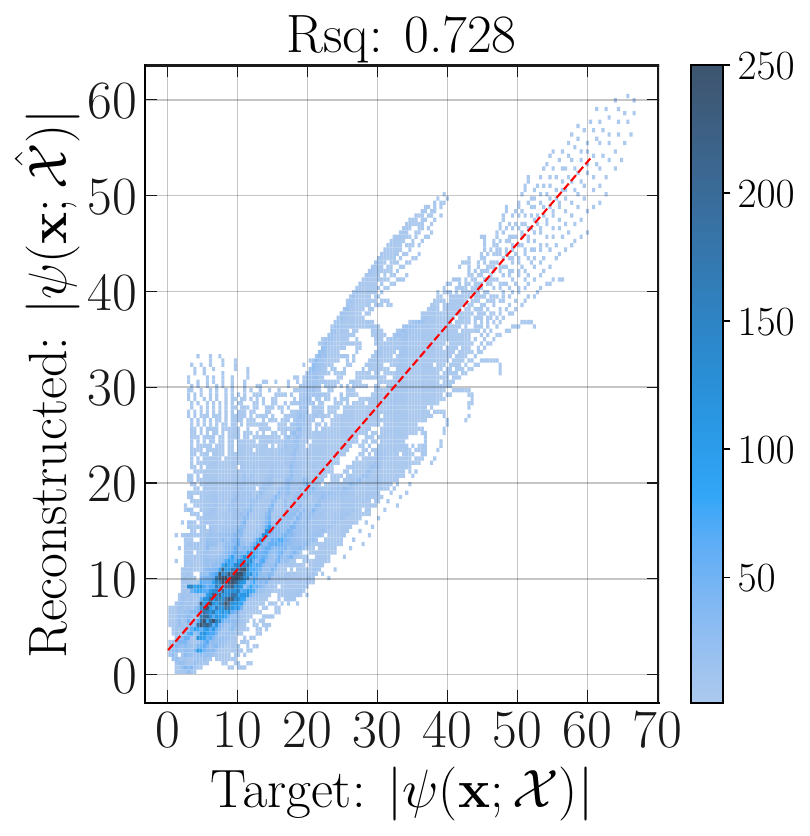}
		\end{subfigure}%
		\begin{subfigure}{.5\linewidth}
			\includegraphics[width=.5\linewidth]{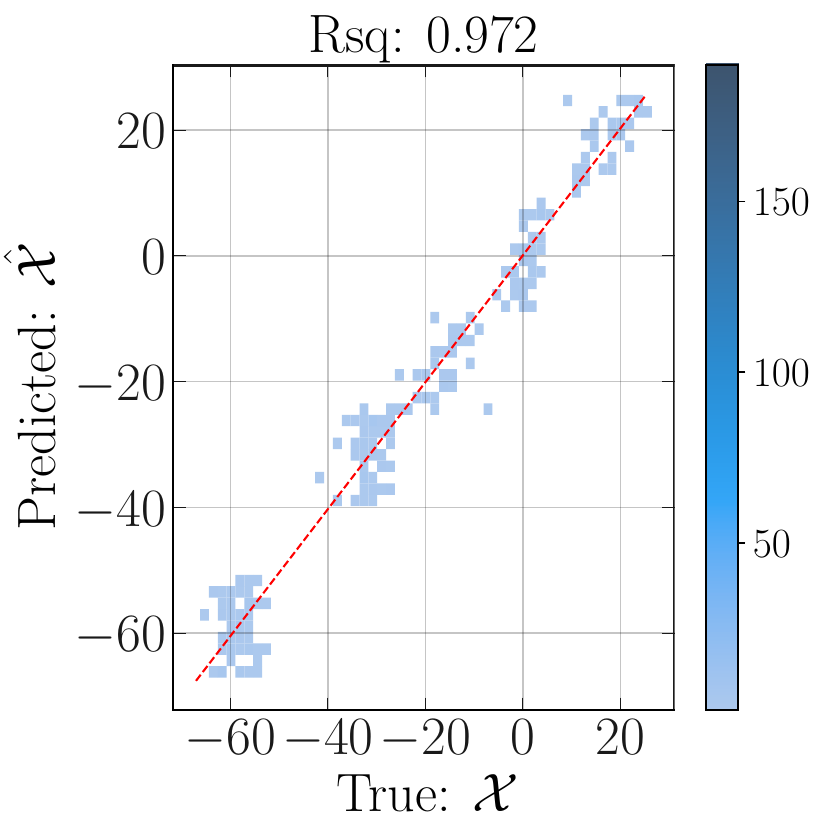}%
			\includegraphics[width=.5\linewidth]{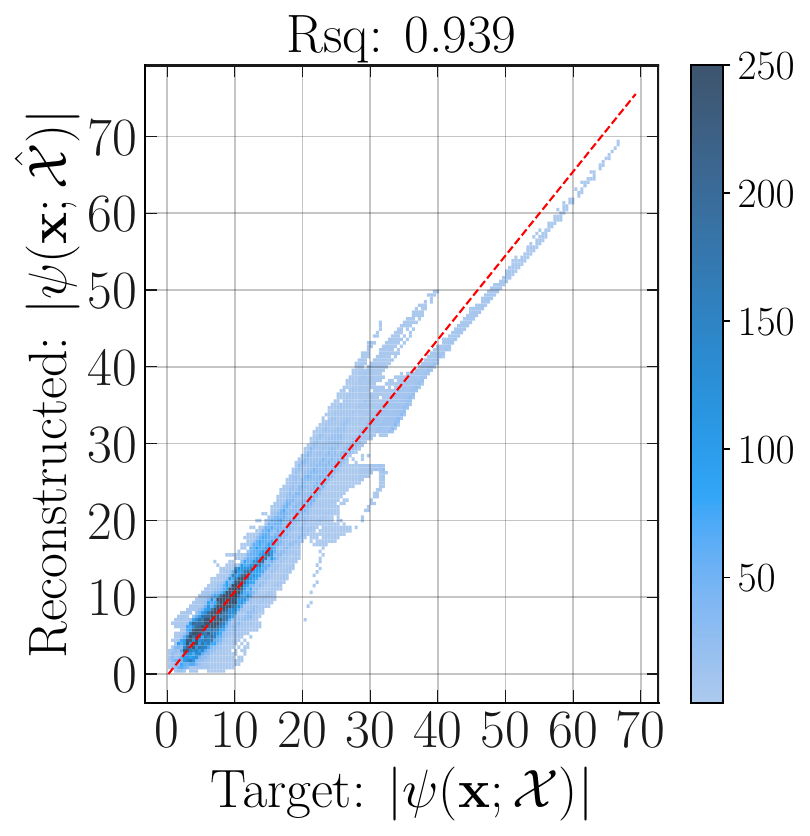}
		\end{subfigure}
		\caption{Downstream problem}
	\end{subfigure}
	\begin{subfigure}{\linewidth}
		\begin{subfigure}{.5\linewidth}
			\includegraphics[width=.5\linewidth]{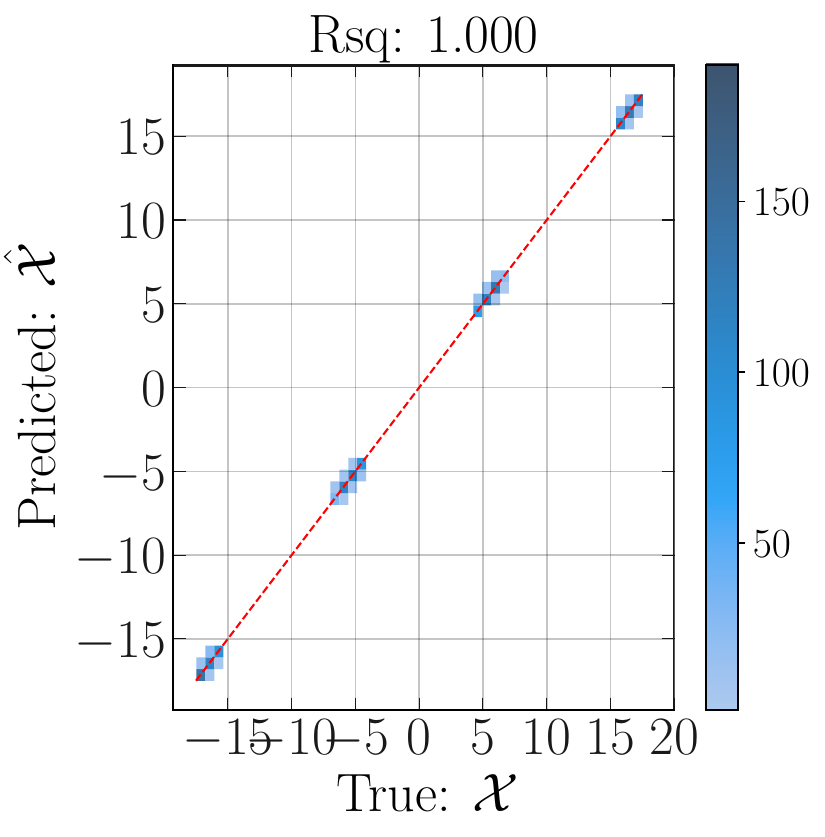}%
			\includegraphics[width=.5\linewidth]{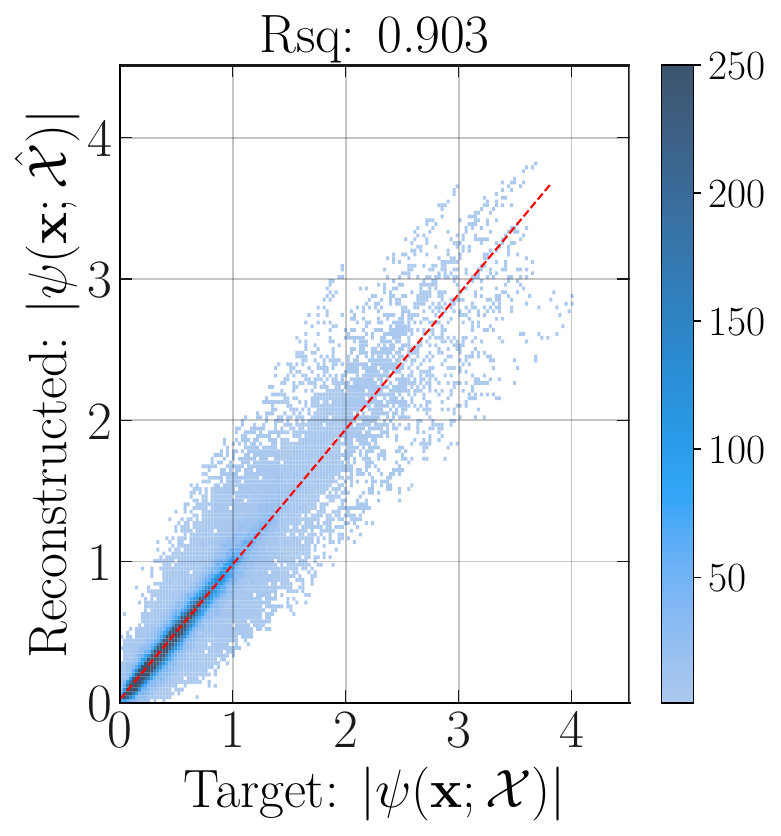}
		\end{subfigure}%
		\begin{subfigure}{.5\linewidth}
			\includegraphics[width=.5\linewidth]{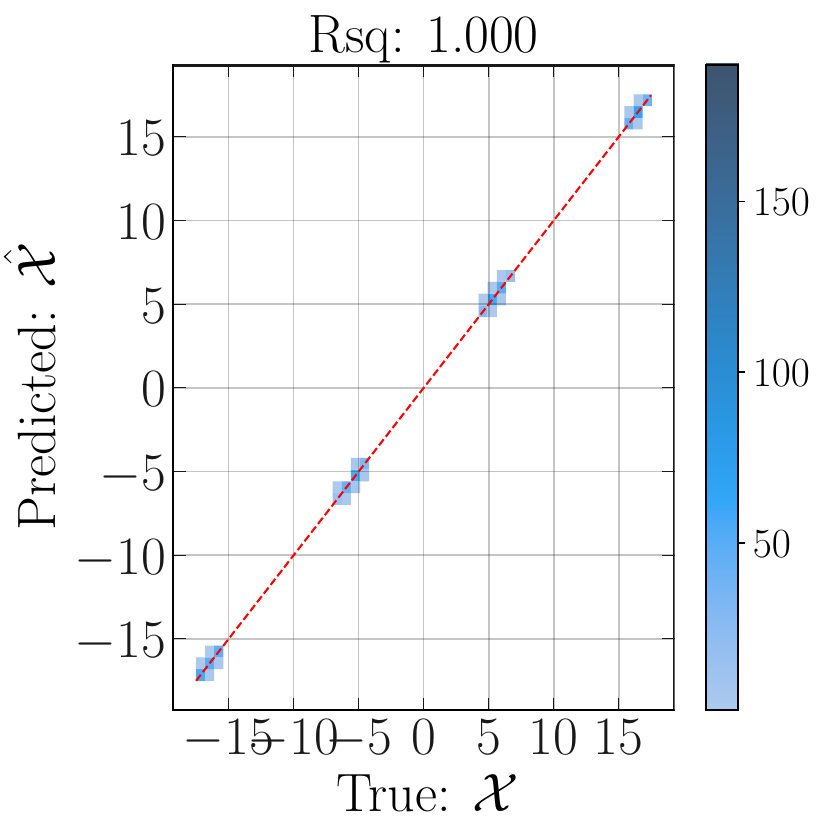}%
			\includegraphics[width=.5\linewidth]{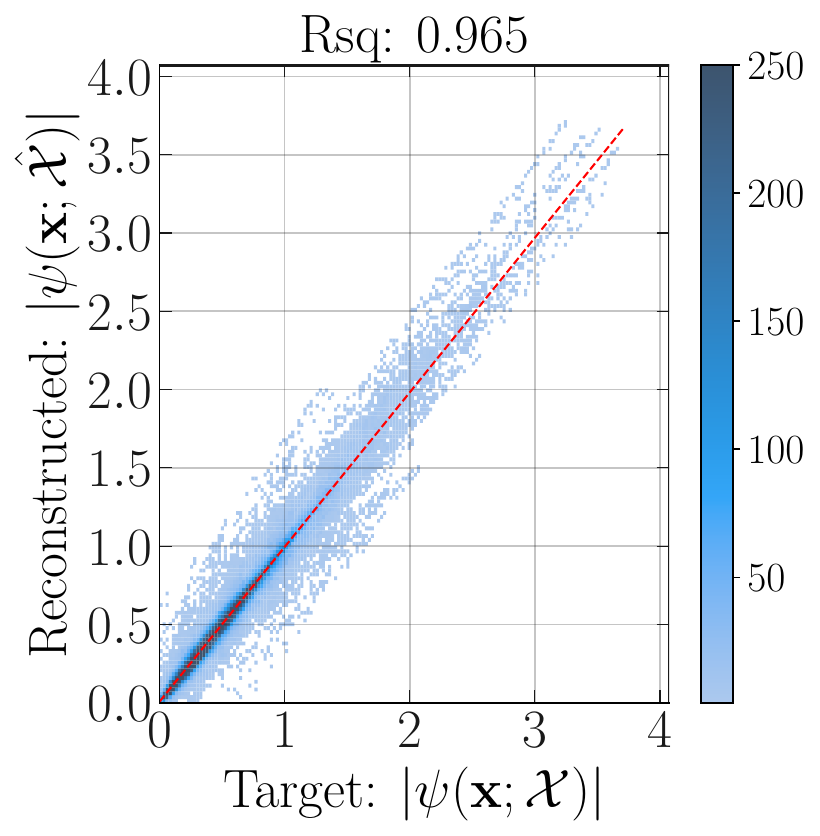}
		\end{subfigure}%
		\caption{Incident problem}
	\end{subfigure}
	\caption{Comparison of true versus predicted scattering positions ($\scatterers$ versus $\hat{\scatterers}$) and the corresponding comparison of true wave amplitude field $|\psi(\var;\scatterers)|$ versus reconstructed wave amplitude field  $|\psi(\var;\hat{\scatterers})|$
		for 30 samples of testing data for  stage I training and stage II training  for the (a) near/far-field problem, (b) downstream problem, and (c) incident problem.}
	\label{FIG: RsqPlots}
\end{figure}

Here we provide a quantitative assessment of the model performance as a remote-sensing and inverse design tool for the problems outlined in Fig~\ref{FIG: ProbTypes}.
This is done by considering a direct comparison between the predicted scattering cluster locations versus ground truth location ($\hat{\scatterers}$ versus $\scatterers$), as well as between the target and reconstructed wavefields ($\psi(\var;\hat{\scatterers})$ versus $\psi(\var;\scatterers)$) for 30 instances of testing data.
We emphasize that the test data was not used in training or validation.
While this straightforward analysis does not reveal the qualitative agreement between the target and inverse solutions, it serves to highlight the performance in the statistical sense and allows for the convenient summary of performance between problem types
We evaluate the models at the end of training stage I as well as at the end of training stage II.
This allows us to assess the importance of the proposed physics-based loss functions, as well as highlight the multi-functional nature of our trained models with respect to the training scheme.

The results of this direct quantitative comparison are depicted by
Fig~\ref{FIG: RsqPlots}.
Superimposed on every plot is a least-squares regression fit of the target-versus-predicted values, with high r-squared coefficients begin recovered for the predicted scattering locations, indicating high effectiveness within the context of remote sensing.
We emphasize that correlation values are recovered for $\scatterers$ versus $\hat{\scatterers}$ is less meaningful in light of the geometric constraints imposed on the scatters; this is especially true for the downstream and incident problems, whereby the scattering arrays are perturbed about nominal grids. However, such metric still allows us to quantitatively compare performance between stages.
It should be noted that in this remote sensing context, Fig~\ref{FIG: RsqPlots}(b) is of the greatest practical interest since the scatterers are not in the viewable wavefields.
Accordingly, this scenario corresponds to the worst performing scatterer prediction model, highlighting the challenge of effectively predicting the scattering arrangements when the immediate effects of the point scatterers are not within $\Omega_{\rm o}$.
High correlation is also achieved for the solution reconstruction ($\psi(\var;\hat{\scatterers})$ versus $\psi(\var;\scatterers)$), as well as significant improvement  across all problem classes when comparing stage I to stage II performance. 
This indicates that the activation of the physics loss is crucial for the models usage as an inverse design tool.
In contrast, however, the scatterer prediction errors ($\hat{\scatterers}$ versus $\scatterers$) do not show much improvement after stage II learning; they are in fact shown to worsen after stage II for the downstream problem (Fig~\ref{FIG: RsqPlots}(b)).

The comparisons in performance between stages I and II depicted in Fig~\ref{FIG: RsqPlots} are consistent with the observation made of Fig~\ref{FIG: Losses}. Namely, that stage I drives down MSE loss of the scatterers, whereas stage II drives down the sparse-reconstruction loss and can in fact elevate MSE loss.
In this regard, we may consider the model trained up to stage I to be sufficient for the purpose of scatterer detection (remote sensing), as it can accurately recover the location of scattering elements for a given a target wavefield. 
However, if the intended purpose of the model is to be used as an inverse design tool capable of providing physically consistent scatterer predictions, then the stage II training and physics-based losses are essential.
This result highlights the non-uniqueness and complexity of multiple scattering solution spaces, as well as the importance of embedding physically relevant penalties when training for inverse design. 


\subsection{Demonstration of the Model}
\label{subsec: Demonstration}
Here we demonstrate the qualitative performance of the model for a sample of test data of each problem type. 
We report no bias in choosing the examples depicted herein.
Rather, these results were picked at random from the generated test data sets and serve to provide a demonstration of the model performing on various scattering fields to complement the quantitative evaluation of the previous section.

\subsubsection{Near/far-field Problem}
\begin{figure}[t!]
	\includegraphics[width=\linewidth]{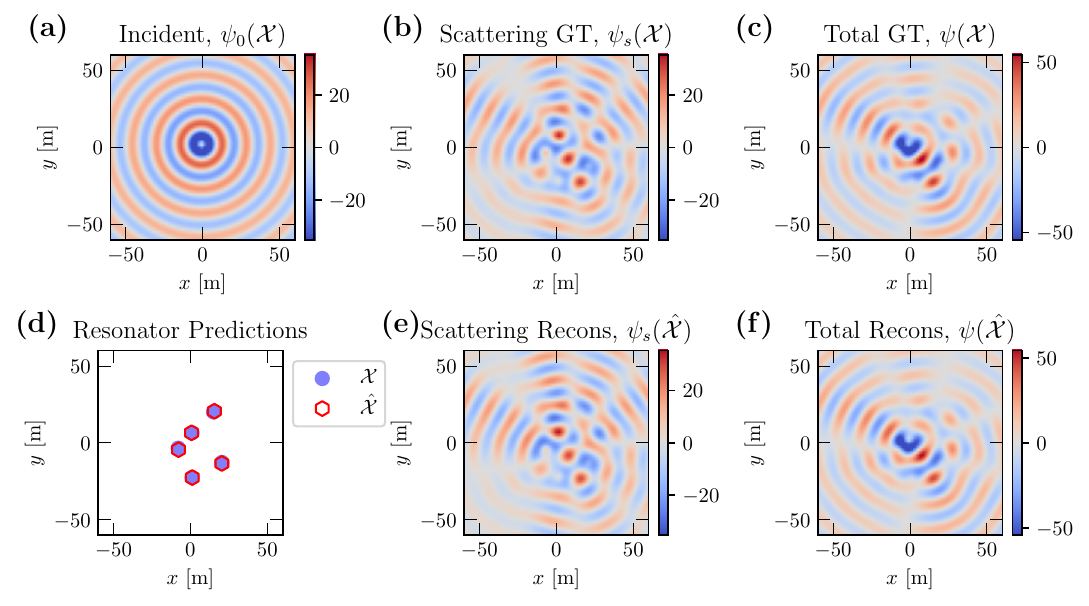}
	\caption{
		Example of the proposed model as an inverse design tool as applied to the near/far-field solutions; (a), (b), and (c) depict the real-valued portion of the incident wave, scattered wavefield, and total wavefield, respectively; (d), (e), and (f) depict the predicted scattering set superimposed with the ground truth scattering set. The scattering solution constructed by evaluating $\psi_{\rm s}(\hat{\scatterers})$, and the total solution given by $\psi_0(\var)+\psi_{\rm s}(\var;\hat{\scatterers})$, respectively.
	}
	\label{FIG: Ex_NearFar}
\end{figure}
\begin{figure}[h!]
	\includegraphics[width=\linewidth]{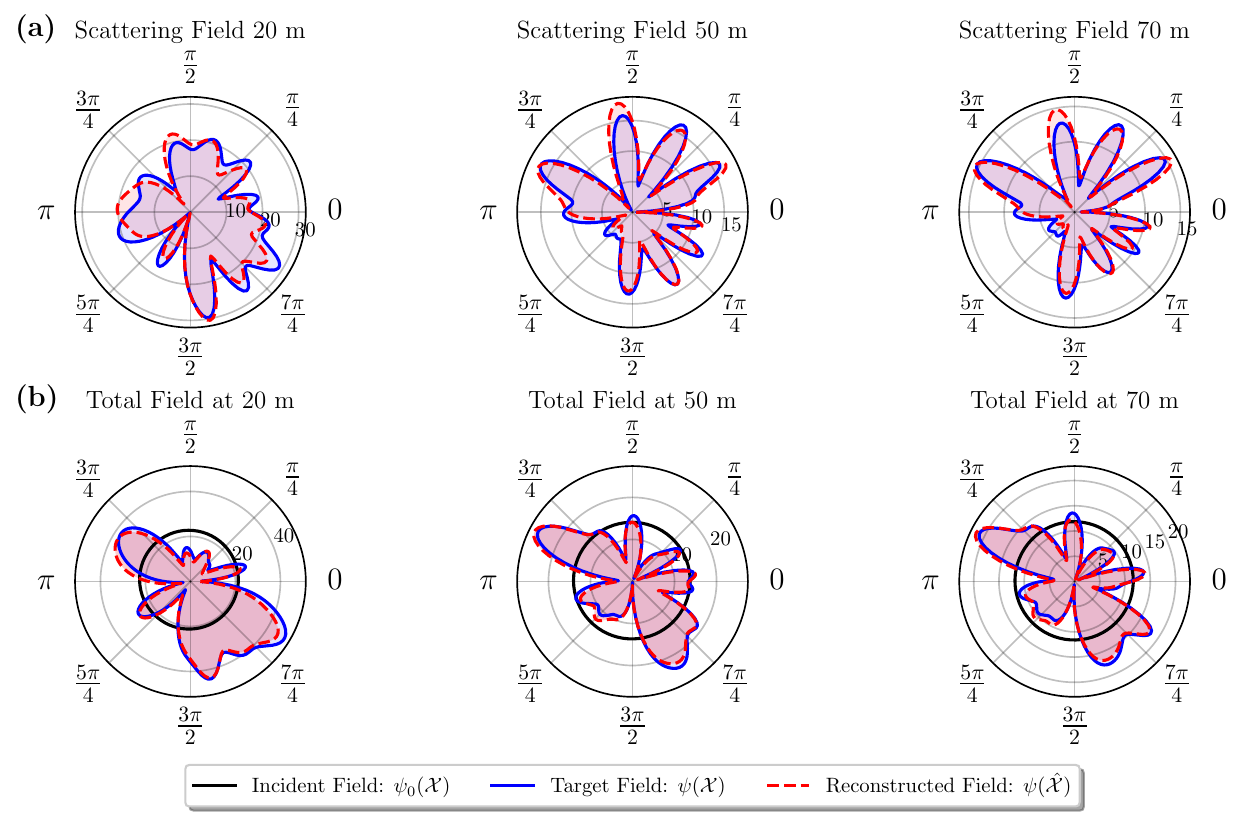}
	\caption{
		Comparison of the solution energies on radial contours $\partial\Omega_{\rm r}$ at radii $r=$ 20, 50, and 70 meters for (a) the scattering solutions and (b) the total solution $\psi = \psi_0+\psi_{\rm}$, with blue depicting solutions generated from the true scattering cluster $\psi(\var;\scatterers)$ and red form the predicted scattering cluster $\psi(\var;\hat{\scatterers})$.
	}
	\label{FIG: Ex_NearFar_rad}
\end{figure}

Figure~\ref{FIG: Ex_NearFar} depicts the evaluation of scattering fields gfor to the near/far-field problem design after stage II training. 
The ground truth solution fields  are given in Figures~\ref{FIG: Ex_NearFar}(a), (b), and (c) for $\psi_0(\var)$, $\psi_{\rm s}(\var;\scatterers)$, and $\psi(\var;\scatterers)$, respectively. 
We reiterate that the model is provided with only $\psi_0(\var)$ and $|\psi(\var;\scatterers)|$, from which predictions regarding $\scatterers$ are rendered.
The predicted $\hat{\scatterers}$ and ground truth $\scatterers$ are depicted in Fig~\ref{FIG: Ex_NearFar}(d), which shows near ideal agreements between the two sets. 
Moreover, the results of the forward model evaluating the predicted scattering clusters ($\sol(\scatterers,\hat{\scatterers})$) are given in Figures~\ref{FIG: Ex_NearFar}(d) and \ref{FIG: Ex_NearFar}(e), which depict the scattering contribution and total wavefield of the reconstructed solution, respectively. Exceptional agreement is recovered with respect to the ground truth fields of Figures~\ref{FIG: Ex_NearFar}(b) and \ref{FIG: Ex_NearFar}(c).

The agreement between the ground truth and reconstructed solution fields is further exemplified by Fig~\ref{FIG: Ex_NearFar_rad} which depicts a polar plot of the solution amplitude at iso-radius contours, $\partial\Omega_r$ (see Fig~\ref{FIG: ProbTypes}).
For both the computed scattering contributions ($\psi_{\rm s}$) and total wavefields ($\psi$), good agreement is recovered between ground truth and reconstructed energy levels (Figs~\ref{FIG: Ex_NearFar_rad}(a)~\ref{FIG: Ex_NearFar_rad}(b)).
Notably, this agreement extends beyond the radius of the observable domain $\Omega_{\rm o}$, indicating the solution agreement extends indefinitely into the far field. 
Hence, the model is shown to work well as a predictive tool both the perspectives of remote detection and inverse design of the near/far-field problem types.

\subsubsection{Downstream Problem}
\begin{figure}[t!]
	\centering
	\includegraphics[width=\linewidth]{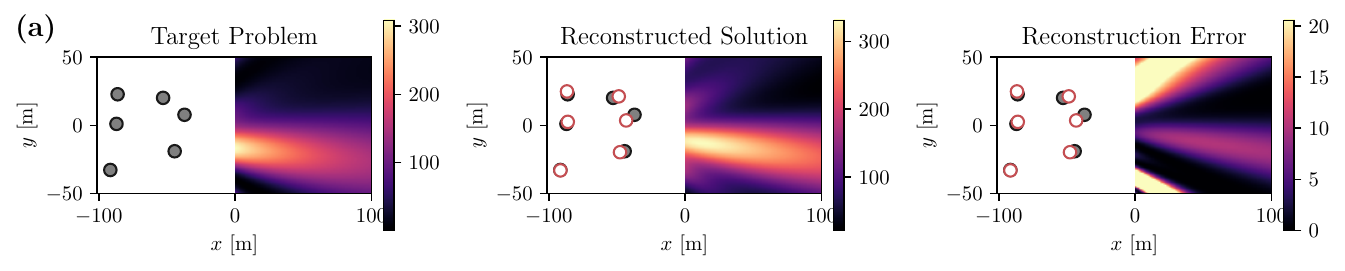}
	\includegraphics[width=\linewidth]{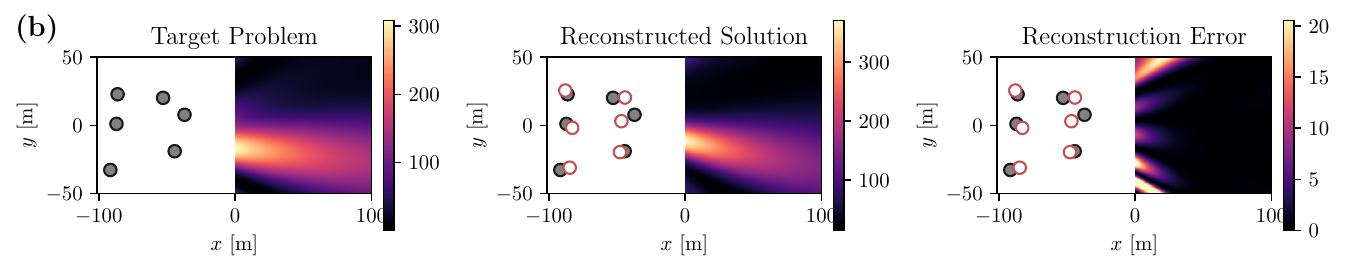}
	\caption{Example result of model performance for a target problem of the downstream problem design for (a) the model after stage I training (no physics-informed losses) and (b) after stage II learning; the stage-I  trained model shows a scattering prediction (red dots) that is in close proximity to true scattering target (gray dots), but that produces a solution field that does not fully capture the true target field; the stage-II trained model predicts a set $\hat{\scatterers}$ that deviates further from the true $\scatterers$, but results in a wave field more consistent with the target wave field. Each subplot depicts the wave in terms of energy, e.g., its squared amplitude.}
	\label{FIG: DS_EX}
\end{figure}

We now consider the model performance when applied to the downstream problem.
Figure~\ref{FIG: DS_EX} depicts the target and predicted scattering clusters, along with the target and reconstructed wave fields, for both stage I and stage II learning, respectively. 
Notably, the model performs seemingly well in the task of remote sensing of the upstream scatterers at the end of stage I, however the reconstructed solutions show considerable error (Fig~\ref{FIG: DS_EX}(a)).
However, when evaluated after stage II (Fig~\ref{FIG: DS_EX}(b)), the model scatterer predictions indeed stray further from the true scattering coordinates as, compared to Fig~\ref{FIG: DS_EX}(a), whereas the reconstruction error is considerably lesser. 
This is indicates that a suitable inverse design was produced for the given target problem after stage II, whereas a closer scatterer detection prediction was produced at the end of stage I.
Hence, Figs~\ref{FIG: DS_EX}(a) and (b) provide a specific example of the quantitative trends reported in Figs~\ref{FIG: Losses} and~\ref{FIG: RsqPlots} with respect to the model functionality at each training stage.

\subsubsection{Incident Problem}
\begin{figure}[t!]
	\centering
	\includegraphics[width=\linewidth]{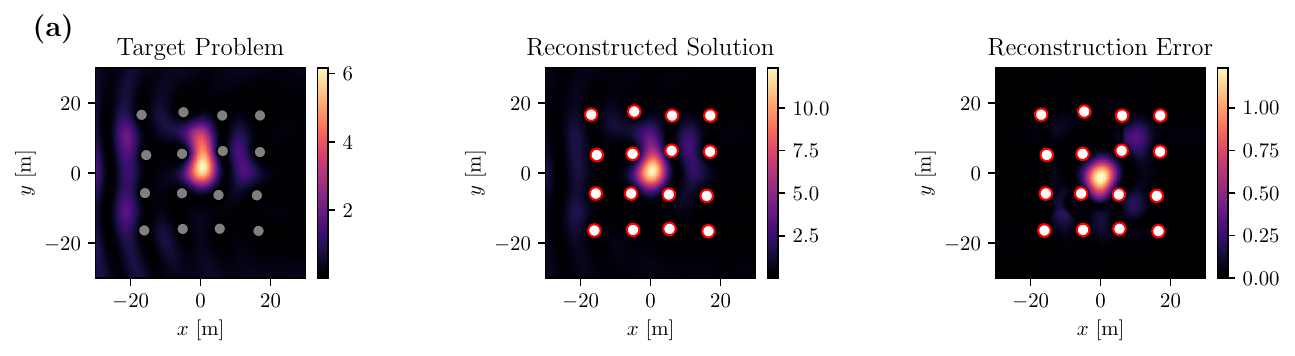}
	\includegraphics[width=\linewidth]{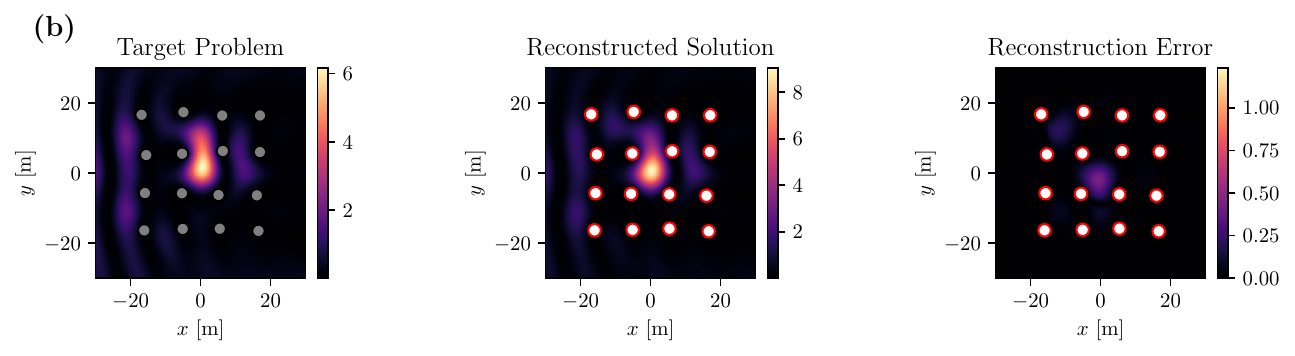}
	\caption{Example result of model performance for a target problem of the incident localization problem design for (a) the model after stage I training (no physics-informed losses) and (b) after stage II learning; while the reconstructed solutions appear similar, the reconstruction errors reveal that the physics-based loss yield predicted scattering cluster that is more consistent with the desired wavefield. Each subplot depicts the wavefield in terms of energy, e.g., its squared amplitude.}
	\label{FIG: Inc_Ex}
\end{figure}

Lastly, we consider model performance of the ML model applied to the incident localization problem.
Figure~\ref{FIG: Inc_Ex} depicts the scatterer array and energy localization patterns for the target and reconstructed wavefields evaluated based on stage I and stage II training.
The targeted wavefield clearly confines energy near the center of the oscillator array; this is despite that fact that the solution field includes the superimposition of the incoming plane wave.
We note that the predicted scattering coordinates align nearly perfectly to their targets for both training stages, with no discernible differences. 
Such good agreement is largely due to the restrictive confinement of the  possible positions of the scatterers in this problem formulation.
However, despite the near identical scatterer predictions, the reconstruction error of the stage II model (Fig~\ref{FIG: Inc_Ex}(b)) performs substantially better within the context of inverse design since the corresponding reconstruction error is considerably lower.
This suggests that while the data-driven losses can accurately place scatterers, the immense complexity and dimensionality of a perturbed localization problem requires the model to either be impeccably precise in its placement of scatterers, or possess insight to the physical relationships allowing for such localization. 
As indicated by Fig~\ref{FIG: Inc_Ex}, it is the incorporation of the physics-based losses that enables such a relationship to be learned.

\subsection{Scattering Field Engineering}
\label{subsec: WaveEngineering}

\begin{figure}[t!]
	\includegraphics[width=\linewidth]{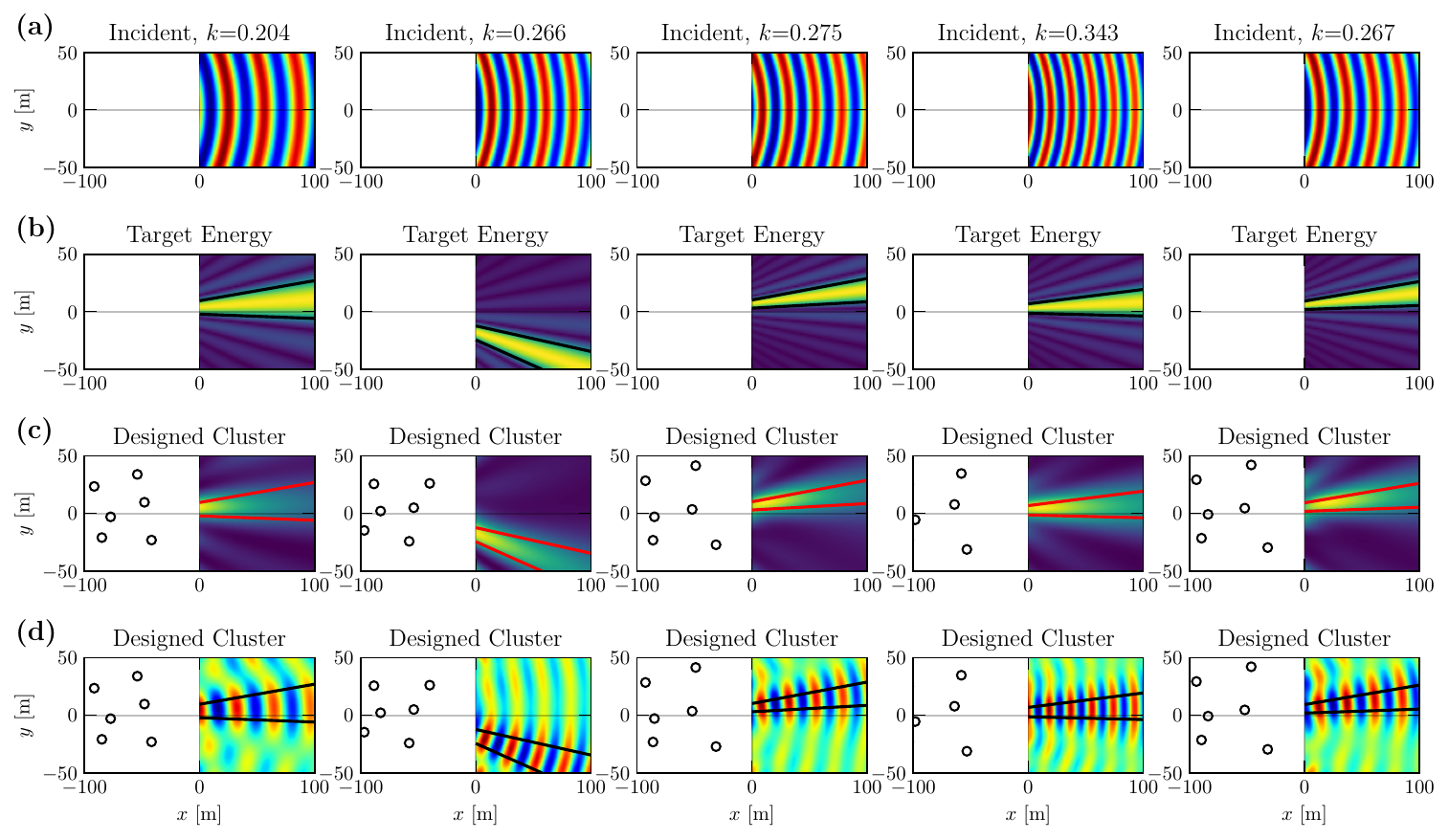}
	\caption{Wavefield engineering of synthetic fields for the downstream problem type: (a) The incident wavefields at variable wavenumbers, (b) the synthetic wavefield constructed to channel energy along a predefined path, (c) and (d) the predicted scattering clusters needed to achieve the desired energy pattern, showing the squared-absolute value and real component of $\psi(\var;\hat{\scatterers})$, respectively.}
	\label{FIG:Synth_DS}
\end{figure}

\begin{figure}[t!]
	\includegraphics[width=\linewidth]{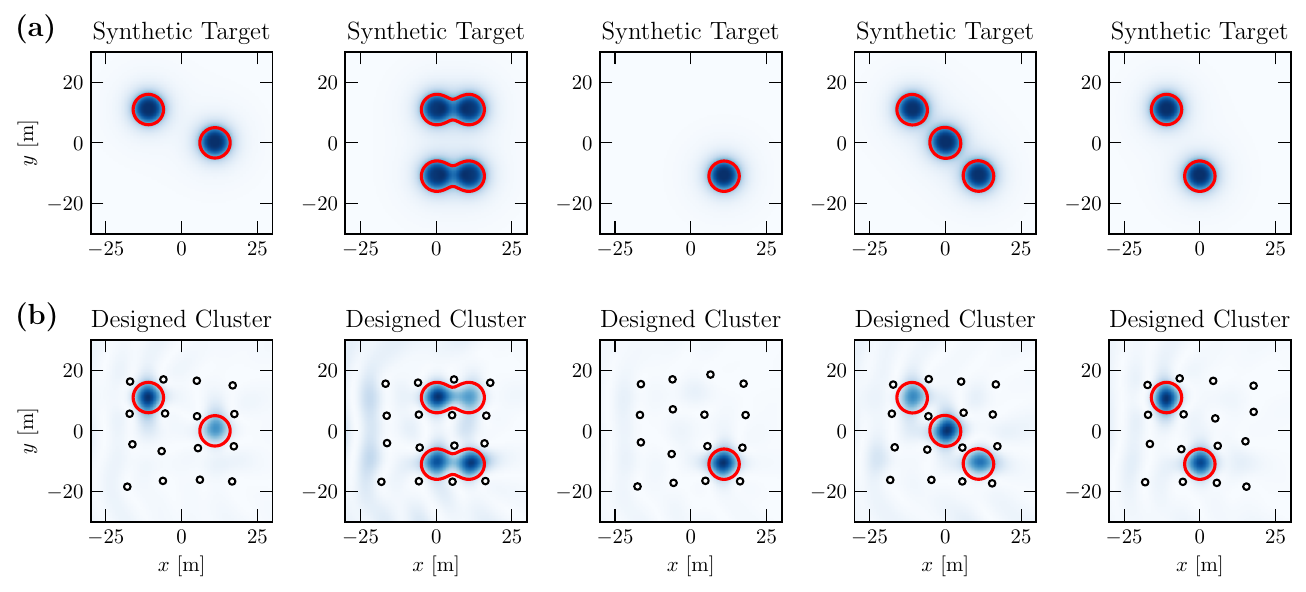}
	\caption{Wavefield engineering of synthetic fields for the incident localization problem: (a) The synthetic wavefield constructed to localize energy at various locations in $\Omega_{\rm s}$, and (b) the predicted scattering clusters needed to achieve the desired energy pattern showing the squared absolute value of the reconstructed wavefield $|\psi(\var;\hat{\scatterers})|^2$, respectively.}
	\label{FIG:Synth_Inc}
\end{figure}
Having demonstrated that the ML model provides effective predictions both in the context of remote detection and inverse design when applied to physical scattering solutions, we now shift attention to the performance of the model on  \textit{synthetically engineering} wavefields.
This challenge carries many practical implications in the context of designing scatterers for a desired functionality, 
since oftentimes a forward solution does not yet exist to emulate the desired performance.
Accordingly, after applying the transfer learning scheme of Section~\ref{subsec: TFL}, several synthetic test wavefields $\overline{\psi}(\var)$ were produced to demonstrate the model's capacity for synthetic wave steering and localization.
In this context, we note that the transfer learning aspect was found to be more necessary for the downstream problem, as the incident solutions did not show dependence on the initialization of stage II optimized model parameters upon the onset of synthetic field learning.

Beginning with the task of directing wave energy along a predefined path as shown in Fig~\ref{FIG:Synth_Targs} (e.g., energy steering), Fig~\ref{FIG:Synth_DS} depicts five testing samples of similar wave pattern design along with the predicted scattering clusters.
The incident wave forms for each desired energy distribution are given in Fig~\ref{FIG:Synth_DS}(a), whereas their target energy fields are depicted in Fig~\ref{FIG:Synth_DS}(b); we note that this problem generalizes across the range of wavenumbers utilized for the downstream problem type.
Figures~\ref{FIG:Synth_DS}(c) and (d) show the model output, e.g., the designed clusters needed to emulate $\overline{\psi}(\var)$, as well as their wavefield solutions $\psi(\var;\hat{\scatterers})$ in terms of the absolute value and real wavefield component, respectively. 
There is a clear agreement between the desired and achieved energy distributions.
Moreover, there is seemingly little dependence on incident wavenumber, meaning that this wave steering is achievable for a wide class of incident waves. 
Hence, in the context of downstream wave steering, we have demonstrated that the proposed model serves as an effective tool for wave scattering engineering.

We next turn to the challenge of localizing energy in a desired subset of the domain (Fig~\ref{FIG:Synth_Targs}(b)); we note that  theses patterns are similar to those of the incident localization problem.
However, we emphasize that the localization observed in the incident scattering fields is a seemingly random function of the scatterer grid perturbations, much like Anderson localization~\cite{Anderson1958}.
However, given that our model was able to learn such nuanced perturbations (Fig~\ref{FIG: Inc_Ex}), it is anticipated that we may employ it to engineer localized patterns into the domain given that the model has learned the underlying relationships causing such localization.
To this end, Fig~\ref{FIG:Synth_Inc} depicts the results for five examples of localization patterns, as well as the designed scattering clusters that the model produced to approximate such energy distributions. 
The generated clusters of Fig~\ref{FIG: Inc_Ex}(b) result in reconstructed wavefields that capture the targeted energy localization patterns of Fig~\ref{FIG: Inc_Ex}(a) remarkably well. 
There are additional ``ripples" of energy in the reconstructed solution field; however, eliminating these altogether is impractical for a real incident wave solution.
Moreover, we comment that the results of Fig~\ref{FIG: Inc_Ex} are bound to the wavenumber $k=\pi/10$, although it is expected that these results would extend equally well to other wavenumbers so long as the nominal grid spacing is scaled accordingly. 
Nevertheless, it is apparent that the model is highly effective in addressing the task of wavefield localization, and therefore, is of great utility for practical engineering applications such as energy harvesting~\cite{Wei2017}.

\subsection{Summary}
Here we provide a brief summary of the results presented in this section.
The proposed modeling strategy, customized loss modules, and training methodology resulted in ML models capable of effectively interpreting target scattering fields and producing meaningful predictions of scattering clusters. 
With specific reference to the objectives stated in Section~\ref{Sec:S2_Problem}, we may deduce the following:
\begin{itemize}
	\item \textit{Scatterer Detection}: 
	Overall, the developed ML models perform well with respected to the remote sensing problem (e.g., scatterer detection).
	For this application, training stage I is sufficient for developing models capable of predicting the scattering clusters within the near vicinity of target scatterer locations. 
	Moreover, the onset of stage II learning does not lead to drastic improvement, and may even inhibit the prediction of scatterer locations if considered solely in the distance minimization sense (see Figs~\ref{FIG: Losses}(b) and~\ref{FIG: RsqPlots}(b)).
	A significant challenge is posed when the scatterer field is outside of the observation domain $\Omega_{\rm o}$, as these problem types yield the poorest remote detection performance. 
	Nevertheless, given the complexity and high-dimensionality of the scatterer detection problem, the developed models prove to be highly capable of accurate predictions.
	\item \textit{Inverse Design}: The developed ML models performed effective inverse design for each of the problem types considered herein.
	Moreover, stage II training highlighted the importance of incorporating physics-based losses in the context of inverse design, since this training stage emphasized learning physically consistent configurations (e.g., the sparse reconstruction loss) over configurations that optimizes the distance minimization problem (e.g., MSE loss of the scatterers).
	This further demonstrates the immense complexity and non-uniqueness of the inverse design problems, since there are many candidate scattering clusters capable of producing a near identical wavefield to a given target.
	Despite this challenge, the trained models were capable of learning the underling relationships of the scattering problems, due in large to the customized physics-based losses.
	\item \textit{Wavefield Engineering}: In the context of wavefield engineering, we have demonstrated the ability to both direct (steer) and localize energy of incident waves using multiple scattering clusters.
	This was achieved in the context of synthetic wavefield designs which were engineered to achieve the desired wavefield energy distributions. 
	We note that this result could not be achieved without first introducing small sets of synthetic training data to the model in the transfer learning stage.
	However, upon the completion of transfer learning, the models proved capable of producing the needed scattering to achieve highly nontrivial wave steering and localization patterns. This renders the ML models highly attractive for wave scattering engineering.
\end{itemize}

\section{Conclusions}
\label{Sec:S5_Conclustions}

In this work, we have presented a physics-informed deep learning framework for the remote sensing and inverse designing  of wave scattering oscillator clusters attached to thin plates.
A convolutional auto-encoder model was deployed to compress wavefield information into a low-dimensional latent space, and multi-layer perception module rendered predictions of scattering clusters.
Physics-based loss quantities were derived to encourage physically consistent scatterer predictions, and a two-stage training scheme was developed in order to maximize the model's performance in the context of remote sensing and inverse wavefield design.

We verified the efficacy of our approach for several classes of multiple scattering problems and compared the disparity in model performance with respect to the problem design.
While good performance was demonstrated both in the context of remote sensing and inverse design, the model faced the most difficulty when predicting scatterers that were outside of the observable domain.
However, due to the embedding of the multiple-scattering solution methodology in to the network loss, the model was still able to recover physically consistent scattering predictions capable of accurately emulating the target wavefields.
This provides a unique insight into the complexity, dimensionality, and non-uniqueness of multiple scattering solutions, and highlights the effectiveness of the sparse-reconstruction physics-informed loss ($L_{\rm sparse}(\scatterers,\hat{\scatterers})$) when inverse modeling such problems. 
To this end, it was shown that the activation of physics-based losses drastically reduced reconstruction error across all problem classes.

There are several notable features of our approach which differentiates it from previous works on this topic.
In contrast to analytical frameworks for solving the inverse scattering problem~\cite{Packo2019,Capers2023}, the proposed deep-learning models are not optimized for any specific wavefield, but are rather constructed to solve the inverse problem for any wavefield of the problem classes discussed herein.
Moreover, we have demonstrated that the models are multi functional in the sense that they can be optimized for different tasks depending on the stage of training, e.g., stage I training for remote sensing, stage II training for inverse design, and transfer learning for wavefield engineering.
Lastly,
while similar works addressed remote sensing and inverse design question for similar scattering problems from a deep learning perspective~\cite{Nair2023}, our work has differed by giving explicit attention to optimizing the multi-functional nature of our models and focusing on complex engineering wavefield patterns. 
In doing so, we have presented a deep-learning inverse modeling strategy that is unique to this class of problems in terms of its multi-faceted training. 
However, this strategy could be readily adapted to other problem types provided that the appropriate domain constraints and physics-based loss quantities are accessible.

Future work could consider alternative modeling strategies, such as equivariant neural networks, to account for the non-uniqueness of the scattering cluster; this could greatly ease the restrictions of the considered problem designs in terms of scatterer placement.
Moreover, it would be of interest to extend the discussed inverse design formulation to other wave phenomena such as Rayleigh waves, shear waves, or electromagnetic waves.
The robustness to noise should be studied as well; this work has focused only on ideal wavefields recovered from forward models.
Lastly, to fully validate the proposed modeling scheme, verification on experimental multiple scattering fields should be pursued.

	\section*{Acknowledgments}
	This work was supported in part by the National Science Foundation Graduate Research Fellowship Program under Grant No. DGE – 1746047. Any opinions, findings, and conclusions or recommendations expressed in this material are those of the authors and do not necessarily reflect the views of the National Science Foundation.
	This work utilizes resources supported by the National Science Foundation’s Major Research Instrumentation program, grant No.~1725729, as well as the University of Illinois at Urbana-Champaign
	
	\clearpage 
	\newpage
	
\clearpage\newpage

\appendix
\newcommand{\utj}{u_j^\mathrm{t}}
\newcommand{\ii}{\mathrm{i}}
\newcommand{\rp}{r_n^\mathrm{p}}
\newcommand{\dd}{\mathrm{d}}

\newpage

\section{Details on Implementing Custom Loss Modules}
\label{APX: Loss}
This section provides the mathematical details required to implement the complex-valued physics loss functions
of Eqs~\eqref{EQ: ForeceVectorLoss} and~\eqref{EQ: SparseLoss} as customized \textrm{PyTorch} modules capable of handling the forward and backward passes of the model.
We define a custom \texttt{greensfunction} module with forward and backward methods. The forward pass returns the complex-valued components of Eq~\eqref{EQ:Gfun}; given some input $r$, this is:
\begin{equation}
	\bar{g}(r) = iH_0(r) - i H_0(ir)
	\label{EQ:Hankel_g}
\end{equation}
Thus, for a given wavenumber $k$ and distance $\var$, Eq~\eqref{EQ:Gfun} may be evaluated in the model as
\begin{equation}
	G_{\omega}(\var) = \frac{1}{8k^2}\bar{g}(k||\var||) \, .
\end{equation}
We note that Eq~\eqref{EQ:Hankel_g} maps from the real to complex domain, $\bar{g}:\mathbb{R}\to\mathbb{C}$. Since the final loss quantities of Eqs~\eqref{EQ: ForeceVectorLoss} and~\eqref{EQ: SparseLoss} are real-valued, this is  not of concern in forward evaluation. However, since the backward pass required real valued gradients for each layer, the appropriate real-valued derivative of Eq~\eqref{EQ:Hankel_g} must be formulated.
As these operations are generally not complex-differentiable, we utilize Wirtinger derivatives per the \textrm{PyTorch} documentation. For our custom module, the Wirtinger derivative of the real-valued loss $L$ with respect to the real-valued input of $\bar{g}(r)$ is defined as
\begin{equation}
	\frac{\partial L}{\partial r} = \mathrm{Re} \left\{ \left(\frac{\partial L}{\partial \bar{g}}\right)^\ast  \frac{\dd \bar{g}}{\dd r}  \right\} \, .
	\label{eq:PyTorch_grad}
\end{equation}
Herein, $(\Box)^\ast$ denotes the complex conjugate and $\frac{\partial L}{\partial \bar{g}}$ is the back-propagated complex-valued derivative of $L$ with respect to the output of our module. Lastly, the derivative of $\bar{g}(r)$ is simply given by the analytical derivative of the Hankel functions as:
\begin{equation}
	\frac{\dd \bar{g}}{\dd r} = -iH_1(r) - H_1(ir) \, .
	\label{eq:g_grad}
\end{equation}

%
\newpage

\section{Spline Approximations of Green's Function}
\label{APX: Spline}
Presently, direct support for the Hankel functions required to evaluate Eqs~\eqref{EQ: ForeceVectorLoss} and~\eqref{EQ: SparseLoss} is not available in \textrm{PyTorch}. Hence, to maintain GPU compatibility for all training computations, we utilized a spline approximations of $\bar{g}(x)$. 
Herein, we use the notation,
\begin{equation}
 	S[f](\bm{x}), \ \ \bm{x}\in\Omega_{\rm spline}
\end{equation}
to denote a spline approximation of the function $f(\var)$ fitted over the domain $\Omega_{\rm spline}$.
With this, we implement the Green's function in practice as,
\begin{equation}
	G_\omega(\var) = \frac{1}{8k^2}\left(S_r(k\var) +i S_i(k\var)\right)
	\label{EQ:Gfun_Spline}
\end{equation}
where the functions $S_r(\square)$ and $S_i(\square)$ represent the real and imaginary parts of the complex Hankel functions contributing to the Green's function:
\begin{align}
	&S_r(\var) = S\left[{\rm Re}\left\{ \bar{g}\right \}\right](\var)\\
	&S_i(\var) = S\left[{\rm Im}\left\{ \bar{g} \right \}\right](\var)
	\label{EQ:Spline_im}
\end{align}
We note that a similar fits were employed to emulate the backwards pass described by Eq~\eqref{eq:PyTorch_grad}.

\begin{figure}[b!]
	\includegraphics[width=\linewidth]{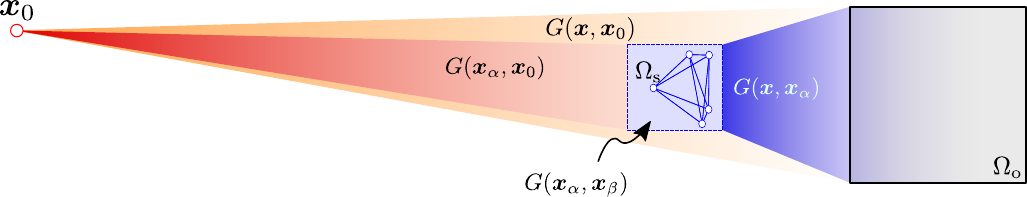}
	\caption{Schematic of the relative distances required for the Green's function to be evaluated for incident forces on the scatterers ($G(\var_\alpha,\var_0)$), incident forces on the observable domain ($G(\var_\alpha,\var)$), inter-scatterer forces ($G(\var_\alpha,\var_\beta)$), and scattering forces on the observable domain ($G(\var_\alpha,\var)$). }
	\label{Fig:Domains}
\end{figure}

The distances considered for incident forces, $G_{\omega}(\var,\var_0)$ and $G_{\omega}(\var_\alpha,\var_0)$, may vary drastically  compared to scattering force $G_{\omega}(\var_\alpha,\var_0)$ and $G_{\omega}(\var_\alpha,\var_\beta)$ depending on the problem configuration (Fig~\ref{Fig:Domains}).
Hence, we fit separate spline functions for incident versus scattering forces so that both may be sufficient well-sampled within their expected range during fitting.
To determine the appropriate fitting domains, we investigate the distances between point loads for 50 test cases of each problem type.
Considering that the the scatters are either upstream (in front of) $\Omega_{\rm s}$ or that $\Omega_{\rm s}\subset\Omega_{\rm o}$, 
we assume that the minimum distance that the incident force must evaluate is the minimum distance from $\var_0$ to $\Omega_{\rm o}$ or $\Omega_{\rm s}$ which may be zero if $\var_0\in\Omega_{\rm s}$. Moreover, the maximum distance is given by the furthest extent of $\Omega_{\rm o}$. 
For scattering forces, the minimum distance is zero since the origin must always be computed along the diagonal of $\textbf{A}$ in Eq~\eqref{EQ: Afb}, and the maximum distance is assumed to be the the far extent of $\Omega_{\rm o}$. Hence, the fitting domains may be mathematically described as,
\begin{align}
	\Omega_{\rm spline}^{\rm incident} = 
	\frac{1}{\lambda}\max \{ 0, \min\{\min\{ D_{0-{\rm o}}, D_{0-\alpha} \} \} \}<x< \lambda\max\{  |\var_0 - \var| : \var\in\Omega_{\rm o}\} \\
	\Omega_{\rm spline}^{\rm scattering} = 0\leq x \leq \lambda \max \{ k|\var_\alpha - \var|  : \var_\alpha\in\Omega_{\rm s}, \var\in\Omega_{\rm o}\}.
\end{align}
where $D_{0-{\rm o}} = \{\min\{\var-\var_0:\var\in\Omega_{\rm o} \}$ is less than 0 if $\var_0\in\Omega_{\rm o}$ and $D_{0-\alpha} = \{\var_0-\var_{\alpha}:\var_{\alpha}\in\Omega_{\rm s} \}$ is less than $D_{0-{\rm o}}$ if the scatterers are to the left of $\Omega_{\rm o}$.
The value $\lambda =1.5$ introduces a region of tolerance to the left and right of the fitting domains.  
To fit each domain, we utilize a 5000 uniformly spaced samples. An additional 1500 samples are included which are logarithimically spaced between $10^{-5}$ and 1 in to enrich the fit near the origin.
Spline approximations were performed using the \texttt{torchcubicspline}\footnote{\href{https://github.com/patrick-kidger/torchcubicspline}{\texttt{https://github.com/patrick-kidger/torchcubicspline}}} library.

\begin{figure}[t!]
	\includegraphics[width=\linewidth]{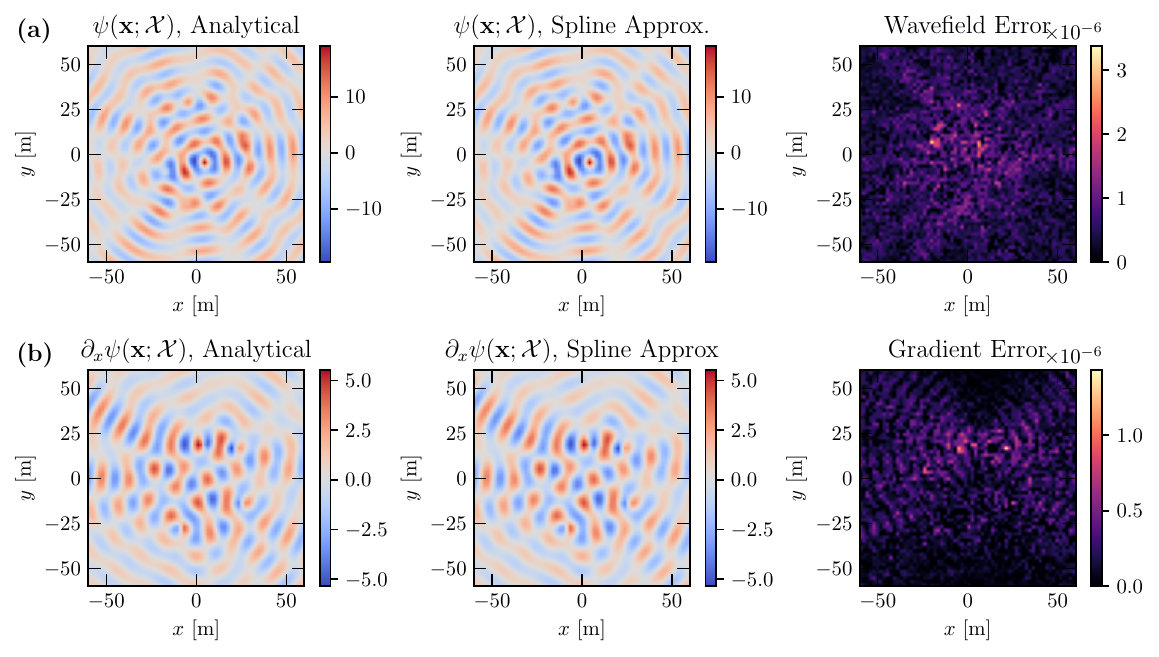}
	\caption{Confirmation of the spline approximations of the Green's function spline approximations: Thee multiple scattering solution $\psi(\var;\scatterers)$ computed with analytical Hankel functions and spline functions, along with their difference, and (b) the same comparison for the function gradients. }
	\label{Fig:SplineRec}
\end{figure}

Figure~\ref{Fig:SplineRec} confirms the convergence of the spine approximations and the true Green's function by considering solution of Eq~\eqref{EQ: MMS_sol}  for a candidate  scattering set $\scatterers$ utilizing both Eqs~\eqref{EQ:Gfun} and  Eq~\eqref{EQ:Spline_im}.
The forward solutions fields of Figure~\ref{Fig:SplineRec}(a) are nearly identical, with the error being many orders of magnitude lower than the solution. The gradients were computed with respect to the scatterers, and equal agreement was found. Hence, we verify that the spline approximations do not alter the physics-based loss function evaluation or backpropagation of the network.

\newpage
\section{Hyperparameter Optimization}
\label{APX: Hyperopt}
Here we provide further technical details regarding the hyperparameter optimization scheme discussed in Section~\ref{subsec:HyperOpt} along with several results from the study.
The hyperparameters of the  model $\mathcal{M}$ are the parameters that are not learnable by backpropagation, but must rather be selected by a practitioner.
These are both architecture-related (kernel widths, layer sizes, etc.) and process related (learning rates, batch size, etc.).
Accordingly, these ad-hoc parameters may  have a drastic effect on the overall model performance. 
Therefore, for the problem at hand, we give special consideration in ensuring that the hyperparameters chosen are most suitable for our goals of scatterer detection and inverse design.
Thus, we  employ a multi-stage hyperparameter optimization routine in order to select the appropriate parameterization of $\mathcal{M}$ for each problem type depicted by Fig~\ref{FIG: ProbTypes}.
 
Our goal is to optimize the model with respect to a given hyperparameter set $\tilde{\bm{\theta}}$.
To do so, we formulate an objective function $J$ describing some performance metric of the hyperparameter set; we note that the structure of the objective function is not known beforehand and must be learned by sampling its parameter space.
In this sense, the objective function is treated as a \textit{black box} which must be learned by making observations. 
It is for this reason that hyperparameter optimization is notoriously costly, since the model $\mathcal{M}$ must be trained and evaluated for each evaluation of the objective function.
Hence, we employ Bayesian optimization which is a common form of active learning designed to optimally sample the parameter space of black-box functions.
In this framework black box function $J$ is emulated by a Gaussian processes which is initialized by a given prior and updated via sampling.
Upon sampling the parameter space, the objective function is modeled as the posterior that maximizes the likelihood of being originated by the observations. 
The primary advantage of Bayesian optimization is its ability to \textit{optimally} sample the parameter space  based on the so-called acquisition function. 
After each sample, a posterior distribution is updated, and the acquisition function is evaluated to determine the next parameter set to observe.
In principle, as the iterations increase, the variance of the posterior should decrease near the predicted optimum. 
In this work, prior distributions were assumed Gaussian with unit variance and zero mean. Posteriors were modeled using the Matern 5/2 Kernel, and the maximum expected improvement acquisition function was utilized.

The hyperparameter optimization of the network presented in Fig~\ref{FIG:  NN_Schematic} may be formulated as,
\begin{equation}
\tilde{\bm{\theta}}^{\star} =	\argmin_{ \tilde{\bm{\theta}}} J(\mathbb{V}|\mathcal{M}( \tilde{\bm{\theta}};\bm{\theta} )),
\label{EQ:BayseOpt}
\end{equation}
where $\mathbb{V}$ denotes the set of validation data and $\mathcal{M}( \tilde{\bm{\theta}};\bm{\theta} )$ represents a candidate model with hyperparameters $\tilde{\bm{\theta}}$ (i.e., learning rate, convolutional kernel size, etc.) and training parameters $\bm{\theta}$. We denote $\tilde{\bm{\theta}}^{\star}$ as the optimal set of hyperparameters for a given objective function.
In total, 16 hyperparameters were identified for optimization (Table~\ref{Tab:Hyper}).
Due to the high-dimensionality of the hyperparameter space, coupled with the high computational cost of evaluating our physics-based losses, it is not practical to evaluate Eq~\eqref{EQ:BayseOpt} over all hyperparameters. 
Therefore, a three stage hyperparameter optimization routine was adopted, with each stage aimed to optimizing a different aspect of the model's performance.

\begin{itemize}
	\item \textit{CAE Optimization --} The first stage of hyperparameter optimization gives explicit focus to determining the best convolutional layers for the CAE. Herein, only the MSE loss of the wavefield reconstruction is considered, and the MLP is turned off altogether. The associated objective function for the Bayesian optimizer is thus:
	\begin{equation}
		J(\mathbb{V}|\mathcal{M}( \tilde{\bm{\theta}};\bm{\theta} )) =  L_{\rm MSE}(\bm{\theta};\mathbb{X},\hat{\mathbb{X}}).
	\end{equation}
	\item \textit{MLP Optimization --} In the second stage, focus turns to optimizing the latent dimension and linear layers of the MLP for best predicting the scatterers postilions. Hence, the MLP is herein activated, however the physics-based losses are not yet turned on. Rather, the focus is given explicitly in parameterizing the fully connected layers; the corresponding  objective function is:
	\begin{equation}
		J(\mathbb{V}|\mathcal{M}( \tilde{\bm{\theta}};\bm{\theta} )) =  L_{\rm MSE}(\bm{\theta};\mathcal{X},\hat{\mathcal{X}}).
	\end{equation}
	\item \textit{Joint Physics/Data Learning Optimization} -- In third stage, focus shifts to optimizing hyperparameters related to the physics-losses and to the overall model learning for both remote sensing and inverse design. These parameters include the loss weights and the learning rates. The training for this stage is performed via the two-stage training routing outlined in Section~\ref{subsec: Training}, and the associated objective function is,
	\begin{equation}
		J(\mathbb{V}|\mathcal{M}( \tilde{\bm{\theta}};\bm{\theta} )) = 
		\tilde{\lambda}_2 L_{\rm MSE}^{\rm I}(\bm{\theta};\mathcal{X},\hat{\mathcal{X}})
		+\tilde{\lambda}_3L_{\rm force}^{\rm I}(\paramvec;\mathcal{X},\hat{\mathcal{X}})
		+\tilde{\lambda}_4L_{\rm sparse}^{\rm II}(\paramvec;\mathcal{X},\hat{\mathcal{X}}),
		\label{EQ:J_joint}
	\end{equation}
	where $\tilde{\lambda}_i$ is a normalized weight and  the superscripts I and II indicate at which stage of training the validation loss is evaluated.
	The weight normalization of Eq~\eqref{EQ:J_joint} was performed by evaluating each  loss for a prescribed 1 meter perturbation from the ground-truth scatterer locations, and scaling their relative contributions to be equal to one-another; their unweighted contributions may differ by several orders of magnitude otherwise. This was done to ensure that no single loss dominates the final stage  of optimization. 
\end{itemize}
The above procedure was applied to each problem type. The hyperparameters that were optimized  at each stage are listed in Table~\ref{Tab:Hyper}. For each stage of hyperparameter optimization, 10,000 data samples were used with a 80-20 split between training and testing data, and 50 iterations of the Bayesian optimizer were executed for each stage. 

\begin{figure}[t!]\centering
	\includegraphics[width=.9\linewidth]{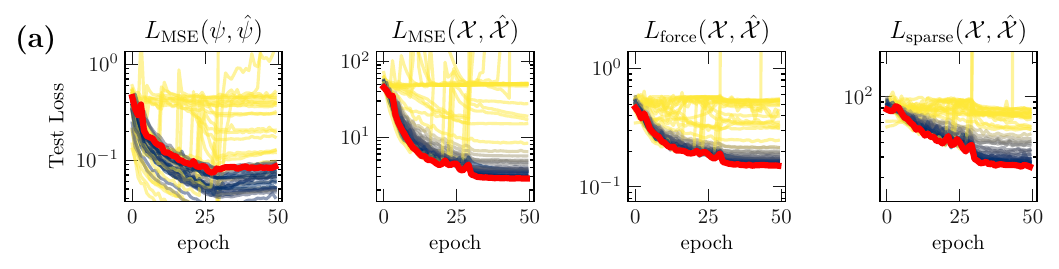}
	\includegraphics[width=.9\linewidth]{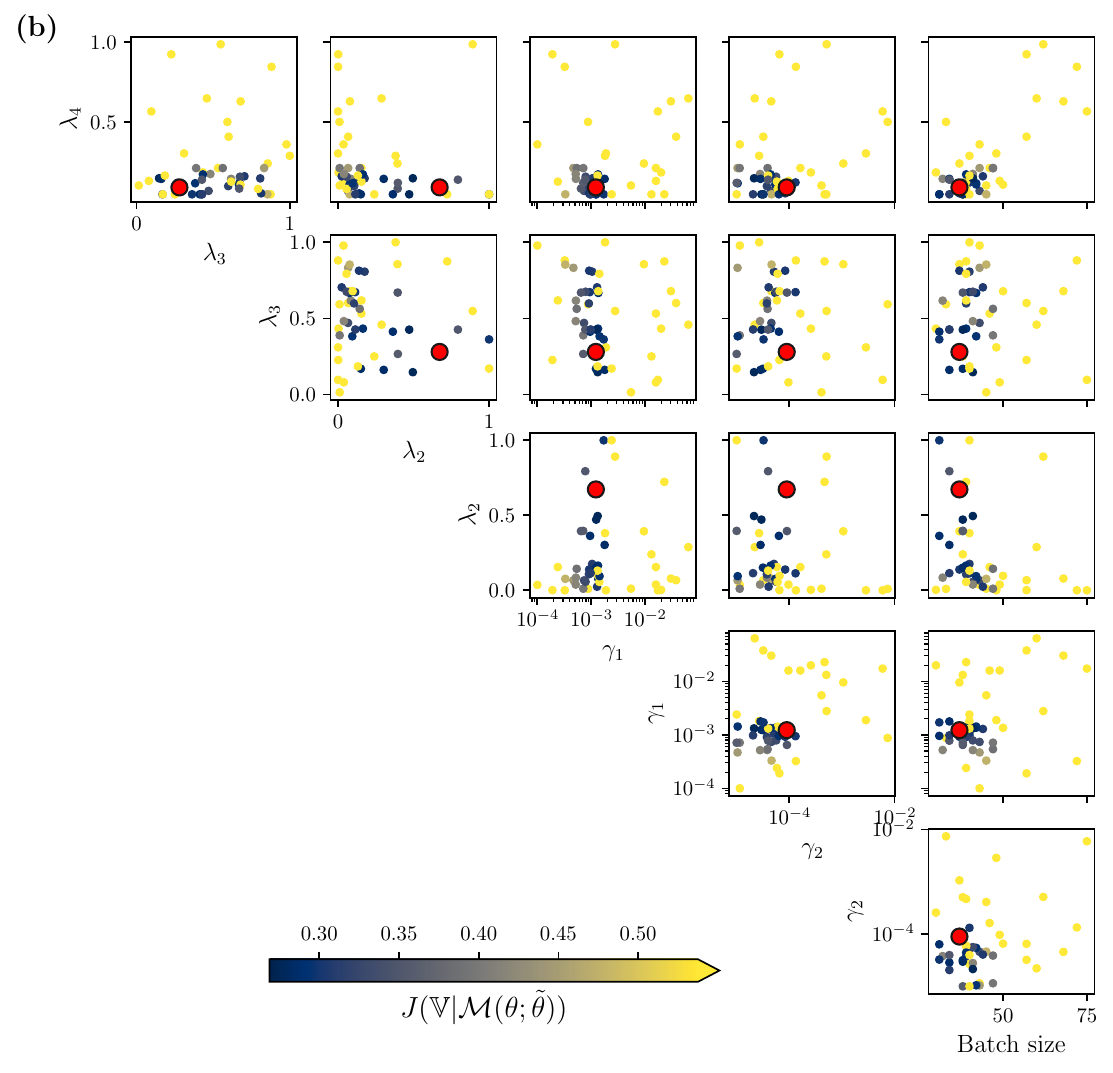}
	\caption{Joint stage hyperparameter optimization results for the near/far-field problem type:
(a) The loss profiles of the four losses considered in Eq~\eqref{EQ: JointLoss}; each optimization trial is superimposed, with the color of the line indicating the value of $J(\mathbb{V}|\mathcal{M}( \tilde{\bm{\theta}};\bm{\theta} ))$, and a thick red line indicating the optimal point. (b) The distribution of sampled parameters with the color indicating objective function value (in correspondence to (a)), and a large red dot indicating the optimal parameter pairing.
}
	\label{FIG:Hyper_near}
\end{figure}

\begin{figure}[t!]\centering
	\includegraphics[width=.9\linewidth]{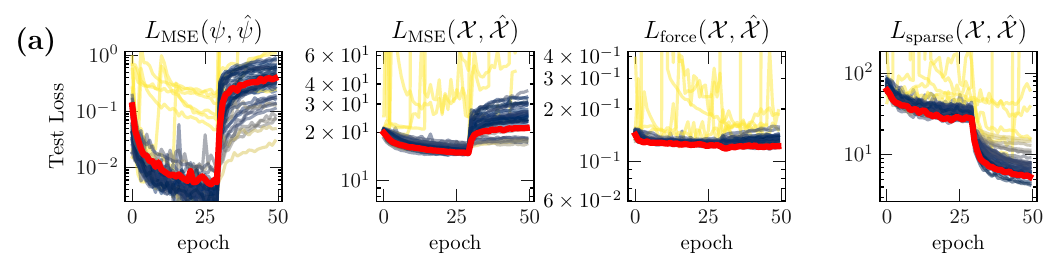}
	\includegraphics[width=.9\linewidth]{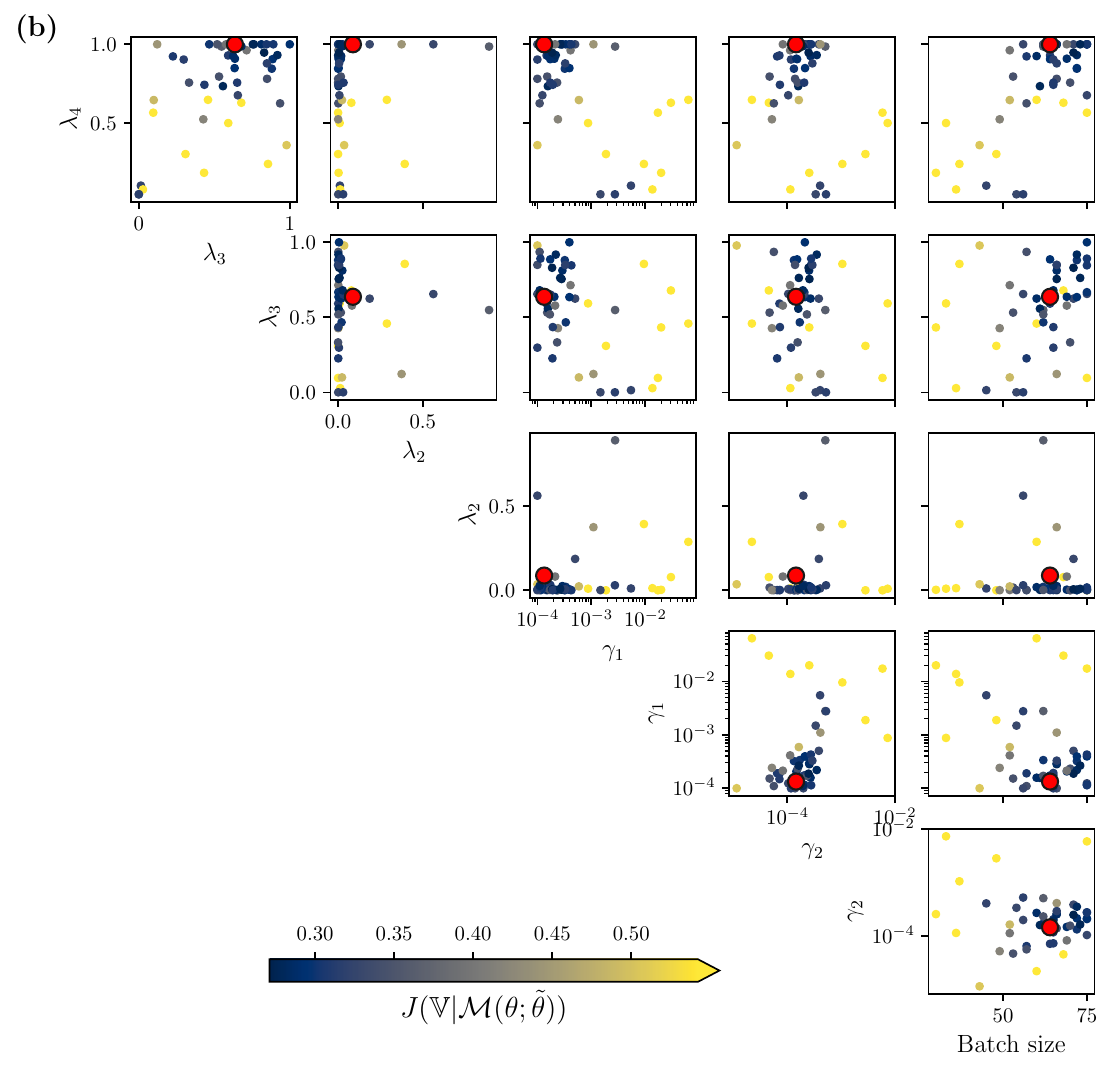}
	\caption{Joint stage hyperparameter optimization results for the downstream problem type:
		(a) The loss profiles of the four losses considered in Eq~\eqref{EQ: JointLoss}; each optimization trial is superimposed with, the color of the line indicating the value of $J(\mathbb{V}|\mathcal{M}( \tilde{\bm{\theta}};\bm{\theta} ))$, and a thick red line indicating the optimal point. (b) The distribution of sampled parameters with the color indicating objective function value (in correspondence to (a)), and a large red dot indicating the optimal parameter pairing.
	}
	\label{FIG:Hyper_DS}
\end{figure}

\begin{figure}[t!]\centering
	\includegraphics[width=.9\linewidth]{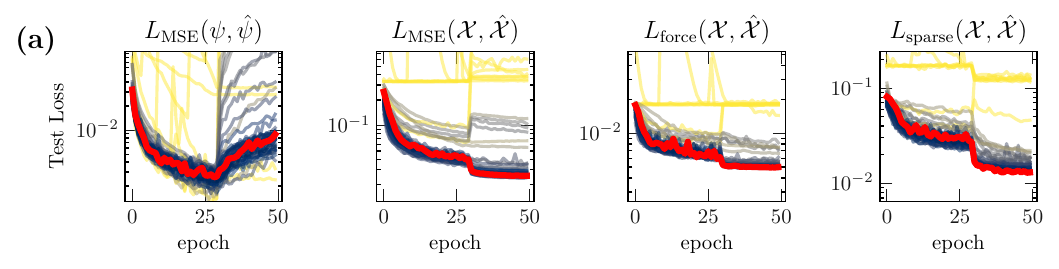}
	\includegraphics[width=.9\linewidth]{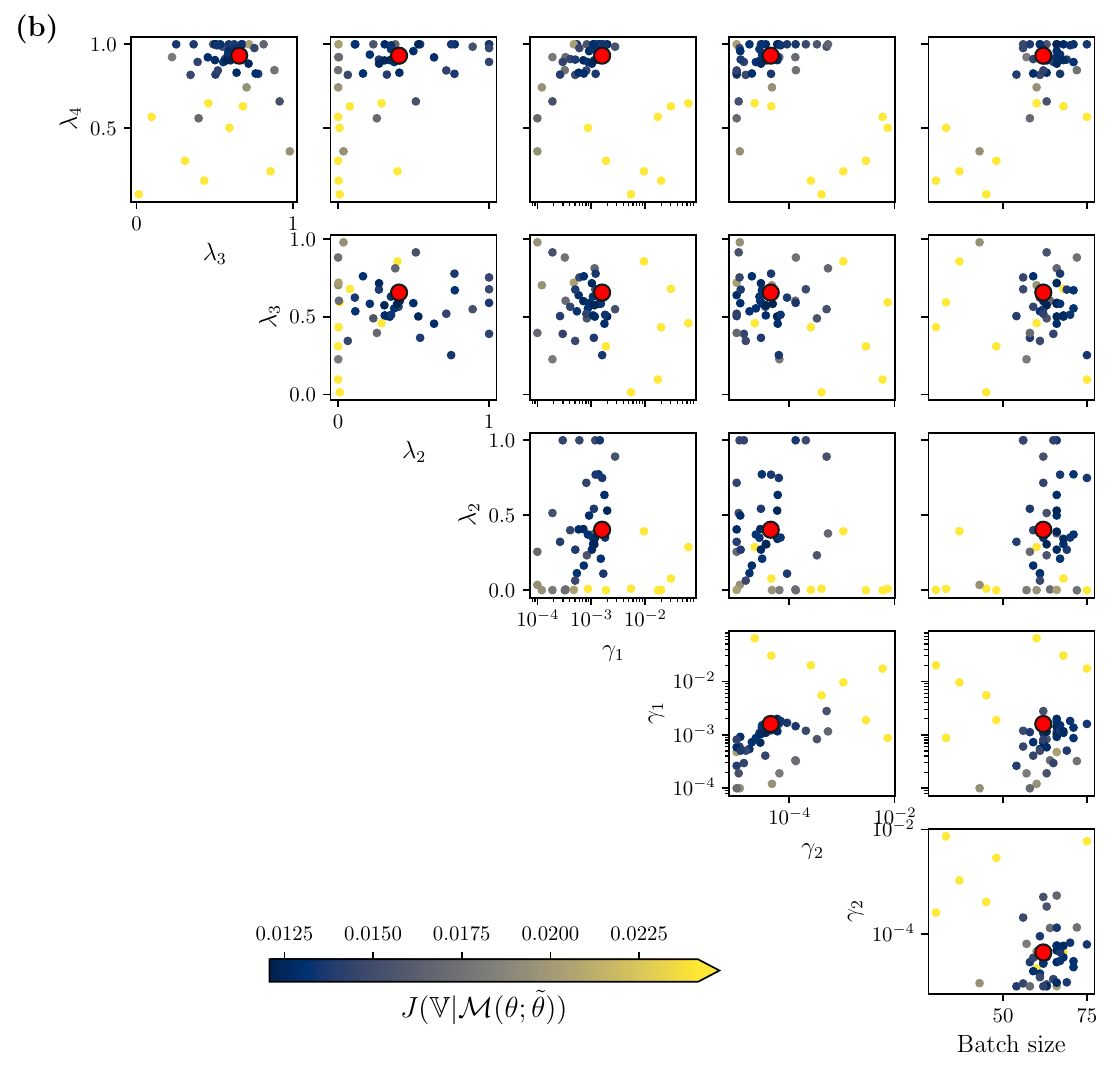}
	\caption{Joint stage hyperparameter optimization results for the incident problem type:
		(a) The loss profiles of the four losses considered in Eq~\eqref{EQ: JointLoss}; each optimization trial is superimposed, with the color of the line indicating the value of $J(\mathbb{V}|\mathcal{M}( \tilde{\bm{\theta}};\bm{\theta} ))$, and a thick red line indicating the optimal point. (b) The distribution of sampled parameters with the color indicating objective function value (in correspondence to (a)), and a large red dot indicating the optimal parameter pairing.
	}
	\label{FIG:Hyper_inc}
\end{figure}

The results of the joint-stage of hyperparameter optimization routine are depicted by Figs~\ref{FIG:Hyper_near},\ref{FIG:Hyper_DS}, and \ref{FIG:Hyper_inc} for near/far-field, downstream, and incident problem types, respectively.
The loss profiles for each of the 50 optimization trials are depicted in  the group of subplots (a) and the corresponding parameter distributions in subplots (b). Optimal parameter selections $\tilde{\bm{\theta}}^{\star}$  are marked by a red dot, and the corresponding  validation losses that minimize $J(\mathbb{V}|\mathcal{M}( \tilde{\bm{\theta}};\bm{\theta} )) $ are denoted with thick red lines.
We note that for each parameter pairing, there is sufficient sampling near the optimal parameter pairing. 
Moreover, the sampled clusters near the optimal value do not deviate significantly in value indicating that a relatively well-behaved parameter distribution was recovered and that the posteriors near the optimum posses low variance.
Moreover, the loss surfaces corresponding to these samples indicate a clear convergence behavior within the context of the objective stated by Eq~\eqref{EQ:J_joint}.
The MSE loss of the CAE is not shown to be minimized by $\tilde{\bm{\theta}}^{\star}$; this is expected as it does not contribute to Eq~\eqref{EQ:J_joint}. 
Recall that we instead seek to optimize the joint loss in terms of the physically-relevant predictions, e.g., the remote sensing and inverse design of the scattering clusters.
We note that in the CAE stage of hyperparameter optimization (not shown herein), it is indeed $L_{\rm MSE}(\sol,\hat{\sol})$ that reaches a minimum; the same is true for $L_{\rm MSE}(\scatterers,\hat{\scatterers})$ at the MLP hyperparameter optimization stage. 
Thus, the hyperparameter optimization strategy posed here is shown to be effective for recovering model parameters that prioritize the modeling objectives of  this work.

\begin{table}[t!]
	\centering
	\caption{Summary of the variables considered for each hyperparameter optimization stage and the selected parameters for (a) the near/far-field, (b) the downstream, and (c) incident problem types.}
	\label{Tab:Hyper}
{\small
	\begin{tabular}{|c|l|l|l|l|l|l|}\hline
	\textbf{Stage }  & \textbf{Parameters} & \textbf{Description }       & \textbf{Bounds}        &(a)&(b)&(c)\\                    \hline
		\multirow{5}{*}{CAE}
		&  $k_i$         					& Kernel size of first layer& [1,6] 					&6&3&3 \\
		&   $k_f$           				& Kernel size of last layer&  [1,6]    					&3&2&2 \\                                      
		&   $l_i$          					& First channel dimension  &  [4,32]   					&32&32&12              \\
		&    $l_f$          				& Final channel Dimension  &  [16,128]   				&125&128& 124             \\
		&     $\sigma_{\rm cv}$         	& Convolutional layer activation & $-$  				&elu&elu&elu     \\ \hline                
		\multirow{5}{*}{MLP}
		&  $n_l$         					& Number of MLP layers & [3,6] 			   				&3&3&3\\
		&   $c_i$           				& First MLP layer   &[124,256]     						&239&216& 123                                     \\
		&   $c_i$          					& Final MLP layer  & [12,124]         					&102&38&12          \\
		&   $\text{dim}(z)$          		& Latent Dimension      &     [12,256]   				&169&256&211  \\
		&   $\sigma_{\rm fc}$         		& Linear layer activation &$-$   						&elu&elu&leaky relu \\ \hline                
		\multirow{6}{*}{Joint}
		&  $\lambda_2$         				& Weight of  $L_{\rm MSE}$& [$10^{-4}$,1]   			&0.67&0.088&0.40 \\
		&  $\lambda_3$           			& Weight of  $L_{\rm force}$ &[$10^{-4}$,1]     		&0.28&0.64&0.64 \\
		&  $\lambda_4$          			& Weight of $L_{\rm sparse}$&    [$10^{-4}$,1]          &0.093&1.0&0.93        \\
		&  $\gamma_1$          				& Stage I learning rate   & [$10^{-5}$,$10^{-2}$]       &0.0012&0.00013& 0.0016       \\
		&  $\gamma_2$         				& Stage II learning rate  &  [$10^{-5}$,$10^{-2}$]   &   8.9$\times10^{-5}$&0.00014&4.4$\times10^{-5}$ \\
	& b$_{n}$ & Batch size for backpropagation & [$15,100$] &32&64& 62\\ \hline           
	\end{tabular}
}
\end{table}



	\clearpage 
	\newpage
	\bibliographystyle{elsarticle-num} 
	\bibliography{MS_ML_BIB.bib}

\end{document}